\documentclass[useAMS,usenatbib]{mn2e}
\usepackage[usenames,dvips]{color}
\usepackage{graphicx,natbib}
\usepackage{multirow}
\usepackage{amssymb}
\usepackage{epsfig}

\newcommand{\chandra}{{\it Chandra}}

\newcommand{\xmm}{{\it XMM-Newton}}
\newcommand{\swift}{{\it Swift}}

\newcommand{\eso}{ESO\,243--49 HLX--1}
\newcommand{\ho}{Holmberg IX X--1}
\newcommand{\ngc}{NGC\,5408 X--1}
\newcommand{\lum}{\thinspace\hbox{$\hbox{ergs}\thinspace\hbox{s}^{-1}$}}

\def\spose#1{\hbox to 0pt{#1\hss}}
\def\laeq{\mathrel{\spose{\lower 3pt\hbox{$\mathchar"218$}} \raise 2.0pt\hbox{$\mathchar"13C$}}}
\def\gaeq{\mathrel{\spose{\lower 3pt\hbox{$\mathchar"218$}} \raise 2.0pt\hbox{$\mathchar"13E$}}}

\title[Long-term X-ray Variability of Ultraluminous X-ray Sources]{Long-term X-ray Variability of Ultraluminous X-ray Sources}
\author[L.~C.-C.~Lin, C.-P. Hu, A.~K.~H.~Kong, D.~C.~C.~Yen, J.~Takata and Y.~Chou]{Lupin Chun-Che Lin$^{1,2}$, Chin-Ping Hu$^3$, Albert~K.~H.~Kong$^1$\thanks{E-mail: akong@phys.nthu.edu.tw}, David Chien-Chang Yen$^4$,
\newauthor
Jumpei Takata$^5$ and Yi Chou$^3$\\
$^1$Institute of Astronomy and Department of Physics, National Tsing Hua University, Hsinchu 30013, Taiwan\\
$^2$Institute of Astronomy and Astrophysics, Academia Sinica, Taipei 10617, Taiwan\\
$^3$Graduate Institute of Astronomy, National Central University, Jhongli 32001, Taiwan\\
$^4$Department of Mathematics, Fu Jen Catholic University, New Taipei City 24205, Taiwan\\
$^5$Department of Physics, University of Hong Kong, Pokfulam Road, Hong Kong}

\begin{document}

\pagerange{\pageref{firstpage}--\pageref{lastpage}}
\maketitle

\label{firstpage}

\begin{abstract} 
Long-term X-ray modulations on timescales from tens to hundreds of days have been widely studied for X-ray binaries located in the Milky Way and the Magellanic Clouds. For other nearby galaxies, only the most luminous X-ray sources can be monitored with dedicated observations. We here present the first systematic study of long-term X-ray variability of four ultraluminous X-ray sources (\eso, \ho, M81 X--6, and \ngc) monitored with \swift. By using various dynamic techniques to analyse their light curves, we find several interesting low-frequency quasi-periodicities.
Although the periodic signals may not represent any stable orbital modulations, these detections reveal that such long-term regular patterns may be related to superorbital periods and structure of the accretion discs. In particular, we show that the outburst recurrence time of \eso\ varies over time and suggest that it may not be the orbital period. Instead, it may be due to some kinds of precession, and the true binary period is expected to be much shorter. 
\end{abstract}


\section{Introduction} 
Over the last two decades, all-sky X-ray monitoring observations of X-ray sources in the Milky Way and the Magellanic Clouds have revolutionised our understanding of X-ray binaries, leading to many important discoveries. 
For instance, detections of X-ray outbursts of transient sources allow us to discover new black hole candidates, and subsequent identification at other wavelengths enable a detailed study of the nature of the system. 
By following the X-ray evolution of X-ray transients, one can study different intensity/spectral states and investigate the nature of the compact object. 
In addition, some persistent X-ray binaries show X-ray variability on timescales from milliseconds to years. 
While observations with high timing resolution can probe variability from milliseconds to hours, orbital periods and superorbital periods (periods longer than the orbital periods) on timescales of days to years can only be studied with regular monitoring observations.

All these observations have been proven to be very successful for our understanding of X-ray binaries, but the sample is mainly limited to the Milky Way which suffers difficulty in determining the distance to the source (hence uncertain luminosity measurements) and high extinction in some regions (e.g., \citealt{naylor1991,thompson2009}). 
Apart from a few cases in the Magellanic Clouds, monitoring observations of X-ray binaries are very rare in other nearby galaxies. 
With the improvement of the spatial resolution and sensitivity of X-ray telescopes, such observations have become feasible now.
One intriguing discovery from X-ray observations of nearby galaxies is ultraluminous X-ray sources (ULXs), which are luminous ($L_{X (0.3 - 10~keV)} > 10^{39}$\lum), non-nuclear X-ray point-like sources with apparent X-ray luminosities above the Eddington limit for a typical stellar-mass ($\sim 10 M_\odot$) black hole.
The majority of ULXs are believed to be accreting objects in binary systems with X-ray flux variability on timescales of hours to years (e.g, \citealt{HVR2009} for short-term variability). 
Assuming an isotropic X-ray emission and that the source does not exceed the Eddington limit, a ULX is the best candidate for an intermediate-mass black hole (IMBH) with a mass of $\sim 10^2-10^4 M_\odot$ (e.g., \citealt{Makishima2000,Miller2004b}).
While ULXs may represent a missing link between stellar-mass black holes and supermassive black holes in galactic centers, their formation and evolution are still not well understood.
Unlike Galactic X-ray binaries, ULXs in nearby galaxies cannot be detected with typical all-sky X-ray monitor instruments because of their X-ray faintness. 
However, one can observe ULXs with large instruments regularly.
In the past 10 years, {\it RXTE}, \chandra, \xmm\ and \swift\ have been used for this purpose. For some well-known ULXs, long-term time series data have been building up allowing us to study their orbital and superorbital periods. Recently, there are some reports on the possible orbital periods of a few ULXs based on \swift\ monitoring observations (e.g., \citealt{Strohmayer2009b}; \citealt{Lasota2011}). Due to the limited time span, the stability of the period which is one of the important indicators for an orbital period requires verification from more observations. Moreover, due to a non-regular sampling and instrumental artefacts, the results have to be cross-checked with different timing analysis techniques.

In this paper, we report an investigation of long-term X-ray variability of four ULXs (\eso, \ho, M81 X--6, and \ngc) that have been observed regularly with \swift. Although the long-term light curves for some of the selected sources in this paper were studied before, we present a systematic timing analysis using both static and dynamic Fourier power spectra on a much longer timeline of data. We also considered the contributions from white noise and red noise to derive the significance of a signal. Finally, we applied a wavelet analysis and Hilbert-Huang transform to compare with results obtained by Fourier analysis. For the present study we concentrate on the usage of different timing analysis techniques for the study of long-term X-ray variability of ULXs.

\section{Our Sample}
In this study, we select ULXs ($L_X > 10^{39}$ erg s$^{-1}$) that have been continuously monitored by \swift\ for more than 4 years. Because \swift\ is not an all-sky monitoring instrument, the sample is limited by the observations in the archive that were originally proposed for different purposes. In order to perform meaningful timing analyses, we require that each target must have more than 200 data points.
In total, we have four ULXs (\eso, \ho, M81 X--6 and \ngc) in our sample that satisfy our criteria (see Table~\ref{obs} for an observing log). 
In the following subsections, we briefly introduce our targets. 

\subsection{\eso}

The ULX \eso\ is the most luminous ULX reaching a peak X-ray luminosities of $1.1\times10^{42}$\lum\ in 0.2 - 10 keV band \citep{Farrell2009}. 
Assuming an isotropic emission in a super-Eddington state, a conservative minimum black hole mass is estimated to be $\sim 500 M_\odot$, making it a very promising IMBH candidate. 
During the past few years, the source exhibits several X-ray outbursts. The regular separation between each outburst of \eso\ is assumed to be caused by the orbital motion \citep{Lasota2011} until recent reports for a delay of the latest outburst \citep{Godet2013,Kong2015}.

\subsection{\ho}
    
\ho\ is a well-known ULX with an X-ray luminosity of $\sim10^{40}$\lum\ in 0.5-10 keV band. It was first discovered by the {\it Einstein Observatory} \citep{Fabbiano88}, 
and has been observed by all major X-ray observatories throughout the last 20 years \citep{Parola2001}.
Recently, \citet{Kong2010} also used \swift\ to monitor the X-ray evolution of \ho\ and utilised the co-added spectra taken at different luminosity states to study the spectral behaviour of the source. 
The best spectral fits are provided by a dual thermal model with a cool blackbody and a warm disc blackbody. 
This suggests that \ho\ may either be a $10 M_\odot$ black hole accreting at seven times above the Eddington limit or a $100 M_\odot$ maximally rotating black hole accreting at the Eddington limit, and we are observing both the inner regions of the accretion disc and outflows from the compact object.    

\begin{table} 
\caption{Summary of observations used in analysis}\label{obs}
\begin{tabular}{lll} 
\hline \hline Source & Date of observations & Data points 
\\ 
\hline \eso\ & 2009 August -- 2013 December & 251
\\ \ho\ & 2008 December -- 2013 October & 483
\\ M81 X--6 & 2009 April-- 2013 October & 396
\\ \ngc\ & 2008 April -- 2012 August & 371
\\ 
\hline 
\hline 
\end{tabular}
\end{table}
 
\subsection{M81~X--6} 
 
M81~X--6, which is also known as NGC~3031~X-11, is the brightest non-nuclear X-ray source in M81. 
The average X-ray luminosity of the source in 0.3-10~keV band is $\sim 2\times10^{39}$\lum \citep{Swartz2003}. 
It was suggested based on optical colours from {\it Hubble Space Telescope} observations that the system contains a $23 M_\odot$ O8 V companion \citep{Liu2002}. Using X-ray spectroscopy, the black hole mass is estimated to be $18 M_\odot$ \citep{Swartz2003}.
Based on the mass of the secondary star, and by assuming that it fills the Roche lobe,  
 an orbital period of about 1.8 days is estimated \citep{Liu2002}.
Although the cadence of the current \swift\ monitoring programme for ULXs is usually from a few days to a week and such a short orbital period is difficult to be detected, we can test if a longer periodicity exists.

\begin{figure*} \centering
\psfig{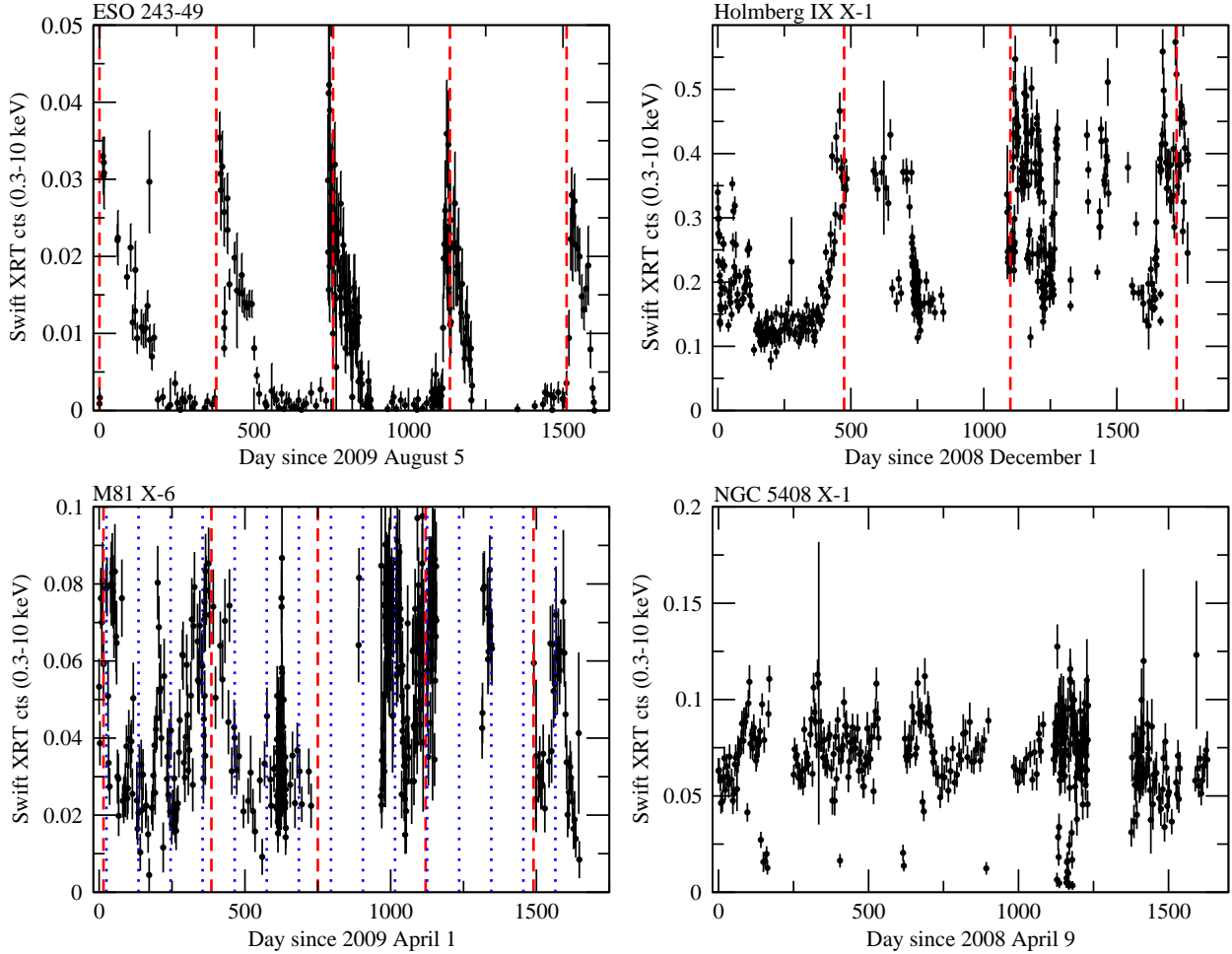}
\caption{\swift\ XRT 0.3--10 keV light curves of \eso, \ho, M81 X--6, and \ngc~ used in our analysis. Data points before the determined starting epoch were eliminated to avoid any fake signals arising from huge data gaps. The red dashed and blue dotted lines are used to label the interval of those intriguing detections summarized in Table~\ref{period}.}  
\label{LC}
\end{figure*}
 
\subsection{\ngc}

\ngc\ is one of the most luminous ULXs with an X-ray luminosity of $\sim 2\times10^{40}$\lum\ in 0.3-10 keV band.
It is also one of the first ULXs observed regularly with \swift. 
Using the \swift/X-Ray Telescope (XRT) data spanning $\sim 500$ days, \citet{Strohmayer2009b} discovered a $115.5\pm 4$-day periodicity and suggested that it is due to orbital modulation. 
If this interpretation is correct, it would support that \ngc\ contains a $\sim1000 M_\odot$ black hole with a $3-5
M_\odot$ companion based on a theoretical simulation \citep{PZ2004}.
\citet{Han2012} also used the weighted wavelet $z$-transform to confirm the periodicity of 115.5 days, but they only considered the light curve within the first 500 days.
However, \citet{Foster2010} argued that such a modulation is due to a superorbital period (see \citealt{Charles2008}) by comparing its physical properties to the micro-quasar, SS 433. 
They suggested that the 115.5-day modulation is originated from a precessing jet, and this scenario implies that the system has a much shorter orbital period and the accreting object is a stellar-mass black hole.
Using 4 years of observations from \swift\ \citep{Grise2013}, an orbital origin of the $\sim 115$-day periodicity can be rejected. 
\citet{Pasham2013} revisited their interpretation as a quasi-sinusoidal X-ray modulation, and re-determined an orbital period of 230 days for \ngc\ based on the average time scale for the recurrence of sharp dips in the X-ray intensity.

\section{Data Reduction and Analysis} 
\subsection{Data Reduction} 
All the data reported in this paper were taken with the X-ray telescope (XRT) onboard \swift. 
We only used data taken in the photon counting mode so that we can identify the precise locations of our targets without any contamination.
We extracted XRT light curves by using the XRT products generator \footnote{http://www.swift.ac.uk/user\_objects} \citep{Evans2007,Evans2009}. It first creates an image from the event list and identifies our target for each observation. 
Only events with energy in the range 0.3--10 keV with grades 0--12 are included. 
A circular source extraction region is chosen to match the point-spread-function.
For the background, an annulus centered on the source is used. 
Then for each observation, source and background counts are extracted. 
Source count rates are corrected for the good time interval, and losses due to bad pixels and bad columns.

Scatter data points can be present at the beginning of each \swift\ monitoring. 
These data are not taken into consideration in our analysis, in order to avoid spurious effects due to large observational gaps.
We also applied a criterion to reject data with uncertainties of the background larger than $3\sigma$. In addition, we removed those data points with anomalously high or low count rates to avoid contamination by the high background or spurious flaring events.
Some data points with count rate errors larger than their count rate due to very short exposures ($\sim 10-30$~s) were not included in the temporal analysis as well.    
Fig.~\ref{LC} shows the original light curves of all four ULXs in our sample.
Since the window function may generate artificial signals, we also re-binned the data samples every 5 days in our analysis to compare with those detections obtained from the original light curve.     
Because the mean cadence of observations is about 2--10 days (excluding obvious data gaps), we only considered periodicities longer than 10 days (i.e., $< 0.1$~1/day in frequency) in our timing analyses.  

\subsection{Methods} 
In order to look for any modulation, we first performed a timing analysis by using the Lomb-Scargle periodogram (LSP; \citealt{Lomb76}; \citealt{Scargle82}), a modification of the discrete Fourier transform which is generalized to the case of unevenly spacing. 
Our static periodogram/power spectra were generated through the REDFIT\footnote{http://www.ncdc.noaa.gov/paleo/softlib/redfit/redfit.html} software and corrected for the bias arising from unevenly spaced data in the time domain that overestimates the high-frequency power \citep{SS1997,SM2002}. Once period candidates are found, we examine their long-term stability by using a dynamic power spectrum (DPS). 
Finally we cross check the results with a weighted wavelet $z$-transform (WWZ) that has been known as an excellent indicator of signals at the expense of being less sensitive to their powers. 

\begin{figure*} \centering
\psfig{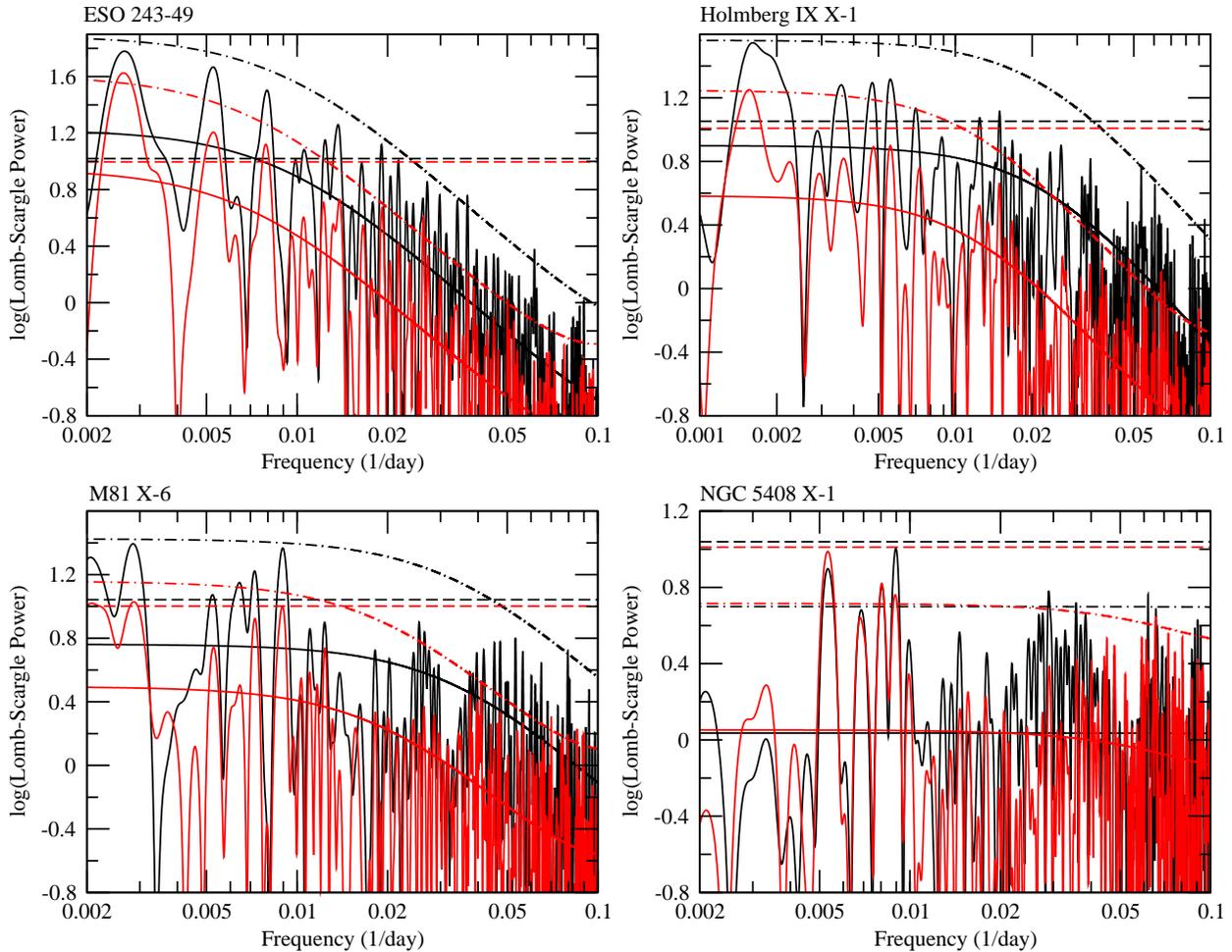}
\caption{Lomb-Scargle periodograms of the four ULXs. The black and the red curves represent the bias-corrected LSPs of the unbinned data and the 5-day average data, respectively. The red noise models for each power spectrum obtained from the unbinned and re-binned light curves are individually plotted as black and red solid lines for comparison. The dashed horizontal and dashed-dotted lines in each panel indicate the 99\% white and red noise significance levels for the unbinned light curves (in black) and the re-binned light curves (in red).} 
\label{LSP}
\end{figure*}

In the following, we will explain the detailed setting of our methods. To derive the LSP using REDFIT, the oversampling (OFAC) and high-frequency (HIFAC) factors, which set the period range and resolution \citep{PR89}, were individually set as 64 and 2 for each process to cover the frequency range in our analysis.
In order to compare with the power determined by a typical LSP, we translated the obtained amplitude of the spectrum into Lomb-Scargle power dividing by a scaling factor related to the average sampling rate and the variance of count rate in our data sets.
The false alarm level of the white noise was set equal to $1-(1-e^{-P})^{N_i}$, where P is the highest peak in a periodogram and $N_i$ is the number of independent trial frequencies. 
In order to simplify our computation, we adopted the empirical function obtained from simulations \citep{HB86} to describe the number of independent frequencies (Horne number; e.g., \citealt{Kong98}). 
We also fit a first-order autoregressive (AR1) process to the time series to estimate the red noise spectrum (solid line in Fig.~\ref{LSP}; \citealt{SM2002}).
The false alarm level of the red noise was assessed through scaling the theoretical red-noise spectrum by an appropriate percentile of the $\chi^2$-probability distribution.   
The analysis with LSP has been widely applied for studying long-term X-ray variability in X-ray binaries (e.g., \citealt{Smale92, Wijnands96, Kong98, Kotze2012}), and REDFIT was also successfully applied to the power spectrum to investigate the red-noise level of long-term X-ray variability \citep{FBS2009}.

For the DPS, we employed a similar method outlined in \citet{Clarkson2003a,Clarkson2003b}. 
The obtained archival data of our targets were analysed with a sliding window to generate the power density, and this window was moved to cover the entire time series. 
This method accounts for variations in the number of data points per interval, while instability of quasi-periodicities can be resolved with a DPS.
The size of the window depends on the length of the periodicity to be examined in the DPS.
The minimum size of the window was set such that it covers at least one complete cycle.
Since suspected periodicities of our detections range from $\sim 30$ to 300 days, the window length was set to 500 days. 
We set the step size of the sliding window in the DPS at $\sim 10$ days to avoid artificial effects. 

The code implementation of WWZ that we used is based on a theoretical study by \citet{Foster96c}.
For an observed time series with $N$ data sets (e.g., $x(t_\alpha)$) taken at $N$ discrete times $\{ t_\alpha:\alpha=1,2,\ldots,N\}$, we can define $N$-dimensional contravariant vector in sampling space with a representation of the canonical basis.
Any function of time $f(t)$ can also be defined as a contravariant vector in sampling space, and therefore the inner product of two functions $f(t)$ and $g(t)$ is defined as 
\begin{eqnarray}\label{eqno1}
<f|g> =\frac{\sum^N_{\alpha=1}w_\alpha f(t_\alpha)g(t_\alpha)}
{\sum^N_{\beta=1}w_\beta},
\end{eqnarray}
where $w_\alpha$ is the statistical weight assigned to uneven data points.
For a set of trial functions $<\Phi_a: a=1,2,\ldots,r>$, the $S$-matrix for the sampling space is given by 
\begin{eqnarray}\label{eqno2}
S_{ab}=<\Phi_a|\Phi_b>
\end{eqnarray}
and the coefficients of equally weighted data can be described as
\begin{eqnarray}\label{eqno3}
y_a = \sum_b S^{-1}_{ab}<\Phi_a| x>
\end{eqnarray}
for the model function.

The wavelet transform (WT) onto the complex trial function at the frequency $\omega$ and the time $\tau$ is
\begin{eqnarray}\label{eqno4}
f(t)=e^{i\omega (t-\tau)-c\omega^2 (t-\tau)^2},
\end{eqnarray}
where $c$ is a constant, and can be regarded as a projection onto the complex trial function of $e^{i\omega(t-\tau)}$ with real statistical weights chosen as $w_\alpha=e^{-c\omega^2 (t_\alpha-\tau)^2}$.
The WWZ consists of the effective number and the weighted variations of uneven data and model functions as below.
According to \citet{Foster96a,Foster96b}, the effective number is  
\begin{eqnarray}\label{Neff}
N_{eff}=\frac{(\sum w_\alpha)^2}{\sum w^2_\alpha}
=\frac{[\sum e^{-c\omega^2 (t_\alpha-\tau)^2}]^2}
{\sum e^{-2c\omega^2 (t_\alpha-\tau)^2}}
\end{eqnarray}
The weighted variation of the uneven data $x$ is given as 
\begin{eqnarray}\label{eqno5}
V_x = \frac{\sum_\alpha w_\alpha x^2(t_\alpha)}{\sum_\lambda w_\alpha}
-\left[ 
\frac{\sum_\alpha w_\alpha x(t_\alpha)^2}{\sum_\lambda w_\alpha}
\right]
=<x|x>-<\mathbf{1}| x>^2
\end{eqnarray}
and the weighted variation of the model function $y$ is 
\begin{eqnarray}\label{eqno6}
V_y =\frac{\sum_\alpha w_\alpha y^2(t_\alpha)}{\sum_\lambda w_\lambda}
-\left[
\frac{\sum_\alpha w_\alpha y(t_\alpha)}{\sum_\lambda w_\alpha}
\right]^2
=<y|y>-<\mathbf{1}| y>^2
\end{eqnarray}
Then the WWZ is read as
\begin{eqnarray}\label{eqno7}
Z=Z(\omega,\tau) = \frac{(N_{eff}-3)V_y}{2(V_x-V_y)}
\end{eqnarray}
Notably, the coefficients $y_a$ are directly computed based on the uneven-smapling data. 
In practice, the choice of $\Phi_\alpha$ could be 
\begin{eqnarray}\label{eqno8}
\Phi_1(t)=1,\quad 
\Phi_2(t)=\cos(\omega(t-\tau)),\quad
\Phi_3(t)=\sin(\omega(t-\tau))
\end{eqnarray}

Once a period candidate is determined from the WWZ, the amplitude of the signal can be determined as WWA = $\sqrt{\sum^r_{\alpha=2} y^2_\alpha}$ \citep{Foster96c}.
All the obtained signals in this paper are subtracted by their mean, and the significance can also be examined with WWA or the corresponding effective number (eq.~\ref{Neff}).  
Here we concentrated on the signals resolved from the evenly sampled data by the WWZ in a comparison to those obtained by the DPS.
Comparing with other time-frequency analysis methods, the WWZ is optimised for non-stationary unevenly time-series analysis, so this method is quite adequate to verify any signal that exists in the obtained data sets. 
 
\section{Results}

\subsection{ESO\,243--49 HLX--1} 

\begin{figure*} \centering
\psfig{file=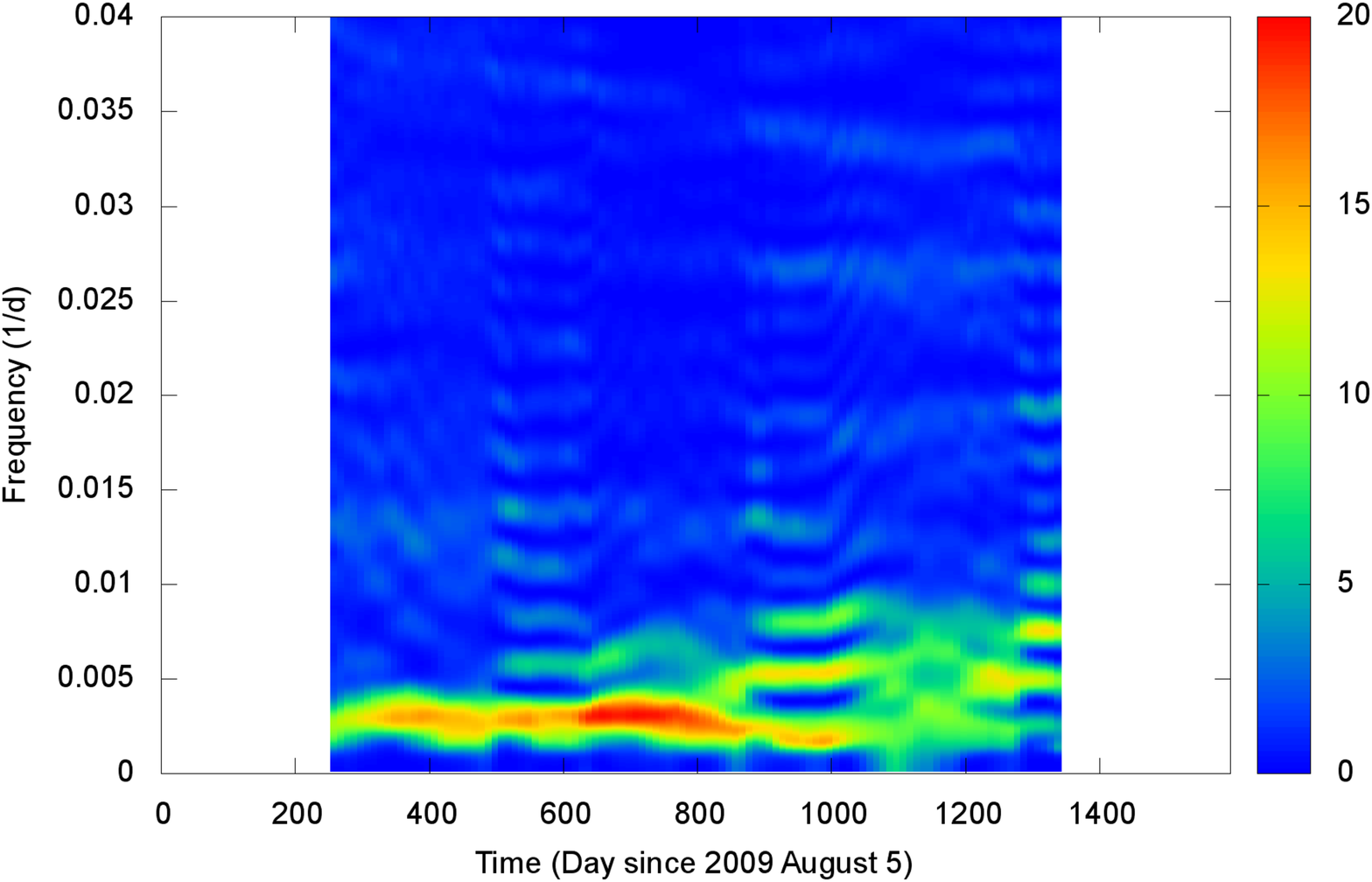,width=3.45in}
\psfig{file=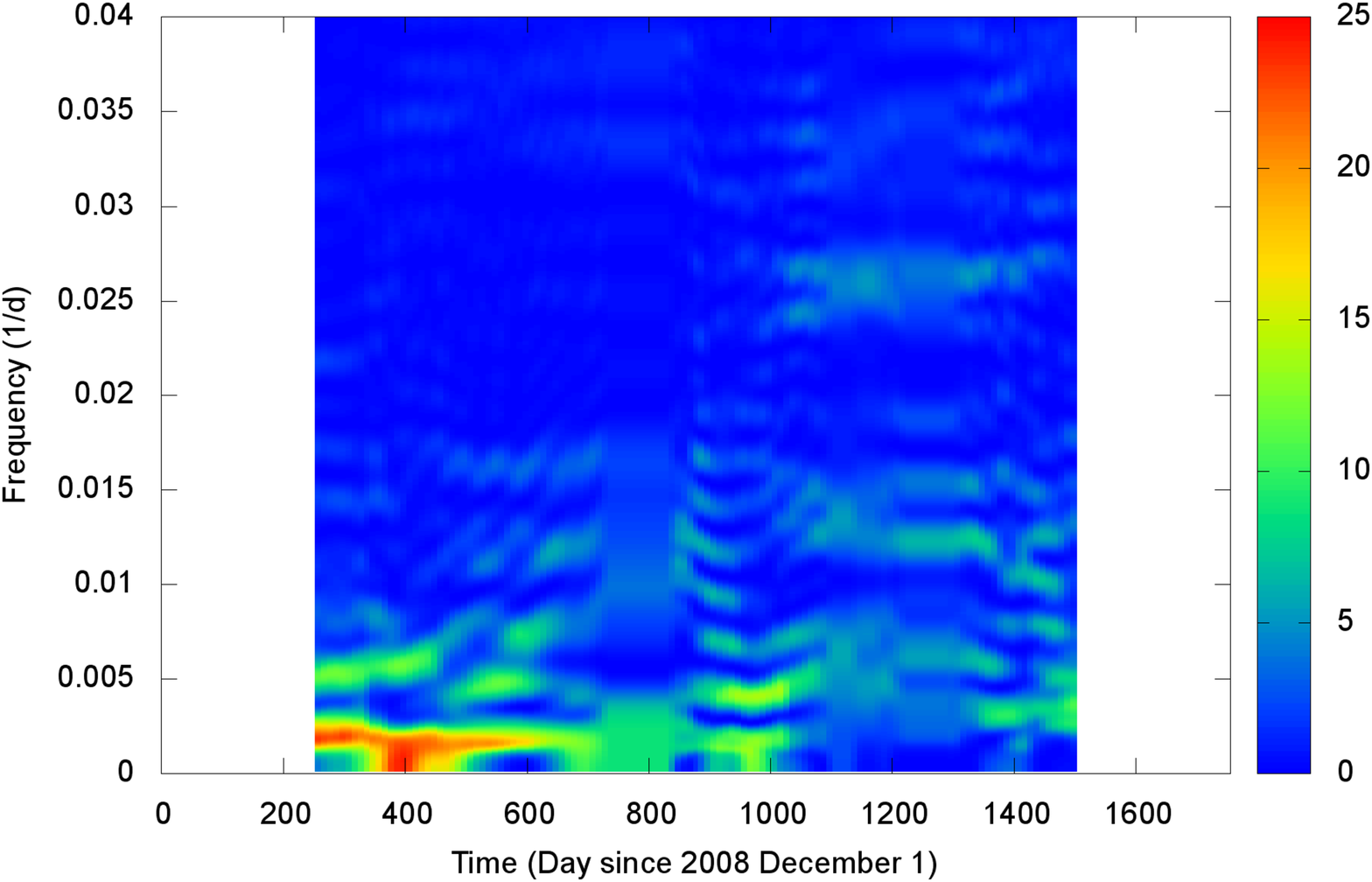,width=3.45in}
\psfig{file=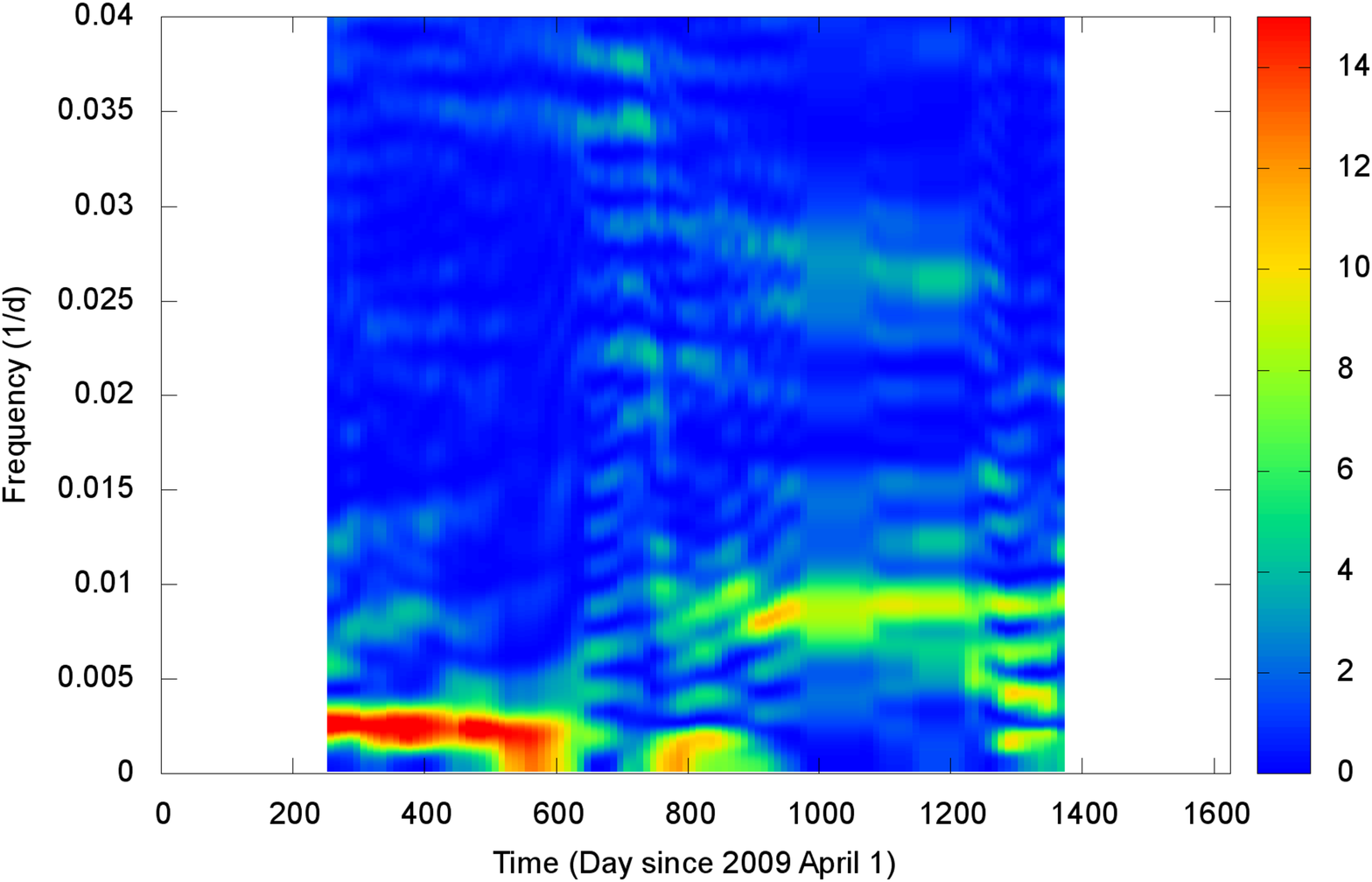,width=3.45in}
\psfig{file=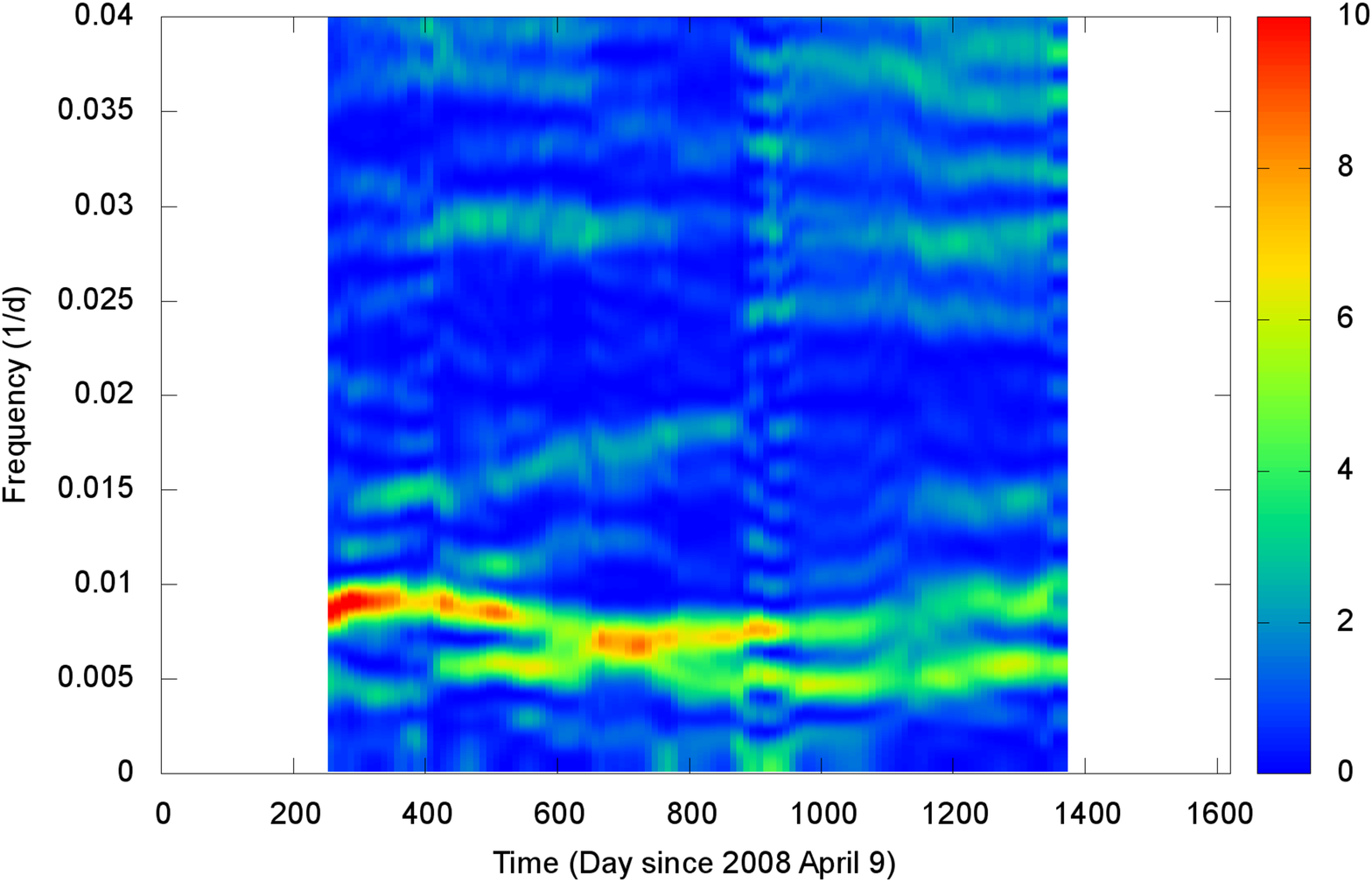,width=3.45in}
\caption{Dynamic power spectra of \eso\ (upper left), \ho\ (upper right), M81 X--6 (lower left), and \ngc\ (lower right). 
The DPSs were obtained with dynamic Lomb-Scargle method with a moving window of 500 days using the re-binned data. The colour of the contour represents the strength of the LSP power.} 
\label{DPS}
\end{figure*}

The long-term light curve of the source is shown in Fig.~\ref{LC}. 
It is clear that the source exhibits substantial variability. 
In particular, there are nearly five complete outbursts with a fast-rise exponential-decay profile similar to typical Galactic X-ray transients. 
The first and second luminosity peaks for \eso\ were in August of 2009 and late-August of 2010 \citep{Godet2010,Kong2010maxi}.
With the regular monitoring by \swift, the next two cycles of outbursts recurred in 2011 and 2012 \citep{Kong2011,Godet2011,Kong2012}.        
The separation between outburst peaks changes from $\sim 350$ days for the first two outbursts to $\sim 370-380$ days for the following ones. 
The most recent outburst was detected in early-October of 2013 with a delay by almost a month compared to the previous one \citep{Godet2013}.

\begin{figure*} \centering
\psfig{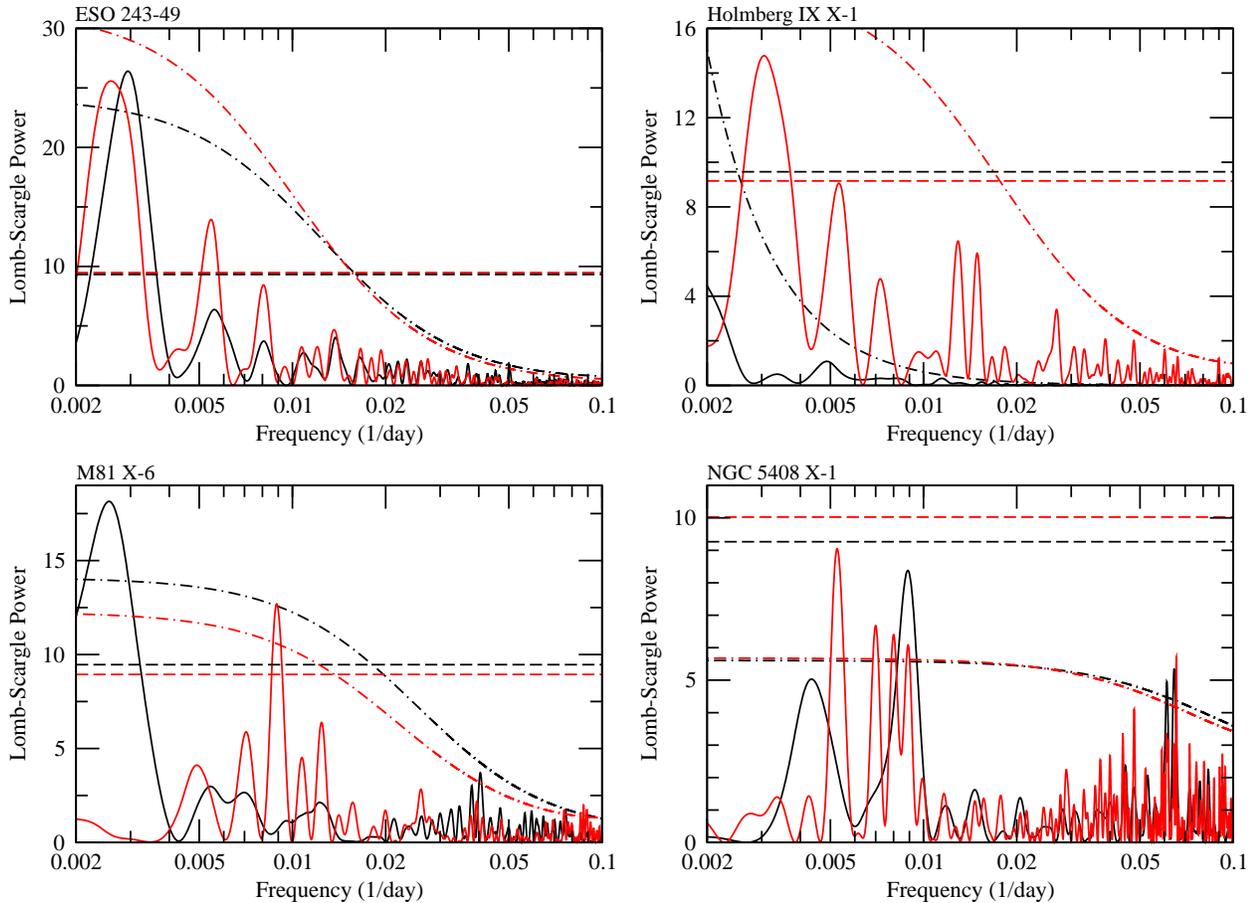}
\caption{Lomb-Scargle periodograms of the four ULXs at different epochs (see Section 4.1--4.4 for the definitions of different epochs). Re-binned light curves were used to derive the LSPs. The power spectrum of each epoch is marked with different colours, while the dashed straight lines and dashed-dotted lines represent the corresponding 99\% white and red noise confidence levels of the LSP power at each epoch. Details of the adopted time intervals are described in the main text.} 
\label{DS_LSP}
\end{figure*}

By performing a LSP analysis, the strongest peak higher than both 99\% red and white noise significance levels for binned data is $378 \pm 6$ days (Fig.~\ref{LSP}), where the error was estimated by propagating two uncertainties.
The first uncertainty in frequency of the trial periodicity was estimated using
\begin{eqnarray}\label{eqno9}
\delta f=\frac{3}{8}\frac{1}{T\sqrt{P}},
\end{eqnarray}
where $T$ is the total time interval and $P$ is the peak power \citep{Levine2011}.  
The second uncertainty accounts for measurement errors effects, and was determined by carrying out Monte Carlo simulations.  
We simulated $10^4$ light curves by randomizing the count rate according to a Gaussian distribution with standard deviation equal to the count rate uncertainties.  
Then, we repeated the LSP analysis and derived a standard deviation for the peak frequency of the periodicity.
The major signal also has a power higher than 95\% confidence level determined from the red noise model for the power spectrum generated from the original data, indicating that this period is statistically significant. 
Similarly, the DPS in Fig.~\ref{DPS} obtained from the re-binned light curve shows a significant periodicity at about 380 days.
There are two other peaks ($190\pm 3$ days and $126.2\pm 1.1$ days) in Fig.~\ref{LSP} with powers more significant than the 99\% white noise confidence level, and they might correspond to different harmonics of the strongest signal as we will describe in detail in section \S 5.1.  
According to previous reports \citep{Kong2011,Godet2011,Kong2012}, the separation between the peak luminosity of the first two outbursts and the following outbursts has increased.
Except for an obvious delay of the outburst occurred in late 2013, because the poor resolution provided by the DPS cannot discriminate such a change unambiguously, we considered the LSP with the sub-divided data sets of day $< 865$ (the black LSP) and day 365-1210 (the red LSP) such that both include three outbursts (see Fig.~\ref{DS_LSP}). 
The strongest signals shown in Fig.~\ref{DS_LSP} are $341\pm 10$ days and $378\pm 13$ days. The difference can be associated with the change in the separation between the two outbursts. 
In addition, we found that the power of the signal at $\sim 190$ days significantly increases in recent observations. 
This result can directly be linked to the second harmonic/first overtone with the separation between recent outbursts ($\sim 380$ days) and provides evidence for variation of the duty cycle for each outburst. 
This can clearly be seen in Fig.~\ref{DPS} as well, and a bifurcation of the main periodicity $\sim 380$ days can be found after day $\sim 815$. 
This might also cause the delay on the most recent outburst in 2013 early-October.
Because of the lack of observations between day 1210--1435, our results were seriously smeared in this time interval.

The aforementioned phenomena discovered in the DPS can also be seen from the WWZ periodogram (see Fig.~\ref{WWZ}).
There are three strong signals, which represent periodicities of $\sim 380$ days, $\sim 190$ days and $\sim 126$ days.
The major signal ($\sim 380$ days) across the whole data is very strong, but the resolution of the WWZ periodogram is not good enough to resolve any variations.
However, we can obviously find that two sub-signals generated from the overtones only get significant after day 720.
This result totally agrees with those from the DPS, indicating a variation of the duty cycle of each outburst.    

\begin{figure*} \centering
\psfig{file=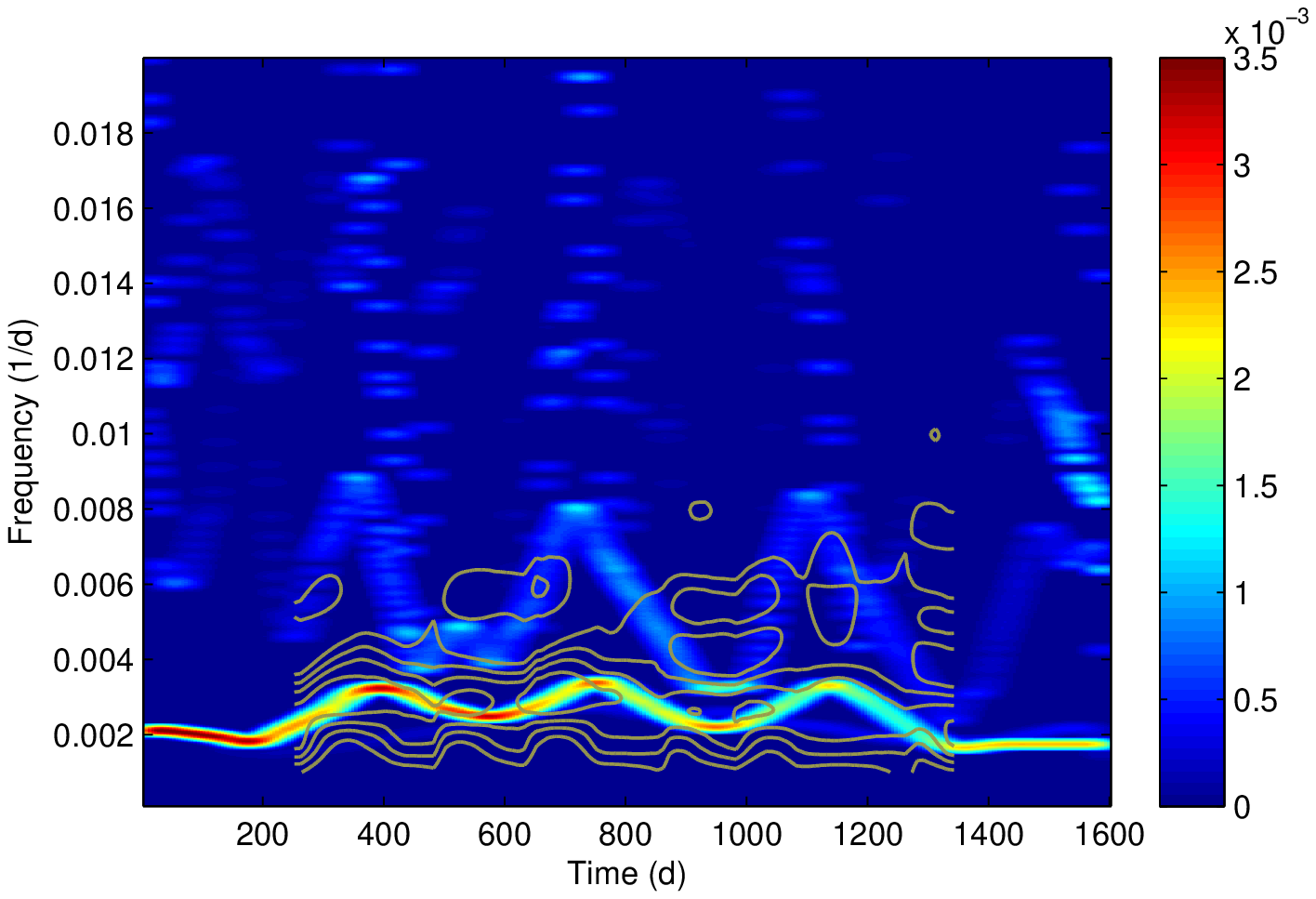,width=6.5in}
\caption{Comparison between DPS and HHT for \eso. The DPS obtained with a moving window of 500 days from the re-binned light curve (contours) is overlapped on the Hilbert spectrum. It is clear that they are consistent with each other.} 
\label{HHT_LSP}
\end{figure*}

Since \eso\ has a regular monitoring before day 1210, it is an ideal object to be further studied with the Hilbert-Huang transform (HHT; \citealt{Huang98}) since HHT analysis requires an evenly sampled data set. 
Though the data were binned with 5 days per bin, there is also an obvious gap after the 4th outburst between day 1210--1435 (only two data points were recorded on days $\sim 1353$ and 1409).
Because this gap is in the quiescent state and the count rate is expected to be less than 0.003 cts/s according to previous observations, we re-interpolated the light curve using a piecewise cubic Hermite spline function \citep{Kreyszig2005} such that the data are evenly sampled without any significant artificial effect.
Following the similar process to study the superorbital period of SMC X-1 with HHT \citep{Hu2011}, we applied an ensemble empirical mode decomposition (EEMD; \citealt{Wu2009}) method to decompose the original light curve into a number of intrinsic mode functions (IMFs) with the medium noise level at first.
The decomposed component of the largest energy/power contains four cycles, which correspond to the main modulation obtained from the LSP and DPS. 
Here, we employed a normalised HHT (Fig.~\ref{HHT_LSP}; \citealt{Huang2003}), which is sensitive to both inter-wave modulation (i.e., the variation of a cycle length) and intra-wave modulation (i.e., the instantaneous frequency changes within one oscillation cycle), to represent the instantaneous frequency on the Hilbert spectrum and to compare with results obtained from the DPS and WWZ. 
A strong variable signal between $\sim$ 250 and 500 days (frequency $\sim 0.002-0.004$~1/day) is seen in the Hilbert spectrum, and it fits well to the major signal yielded from the DPS.
The variation shows roughly five cycles, which is consistent with the quasi-periodicity detected from the light curve. 
Since the $\sim 380$-day modulation is highly non-linear (non-sinusoidal), the intra-wave modulation would dominate the variation of instantaneous frequency.

\subsection{Holmberg IX X--1} 
Fig.~\ref{LC} and Fig.~\ref{LSP} show the long-term light curve and the LSP of \ho, respectively.     
Comparing to \citet{Kong2010}, \ho\ has gone through at least three other high/low intensity transitions. 
The two period candidates ($\sim 28.8$ and 64.5 days; \citealt{Kaaret2009b}) found in previous analyses below the 85\% white/red noise confidence level in the LSP. 
Instead, 
we obtained a significant periodic signal from both unbinned and re-binned data sets at $625 \pm 20$ days, which reaches the 99\% red noise significance and exceeds over the 99\% white noise significance levels in Fig.~\ref{LSP}.
According to the light curve, the peak ($\sim 0.4$ cts/s) recurs at days 440--455, 1120--1130 and 1730--1740, and each separation is about 625 days.
It can also be seen at days $\sim 645$ and 1270--1275 with a separation of $\sim$ 625 days.
If we consider the dip features determined from the light curve, we can also find them with a separation of about 600 days at days 135--140, 750--760, and 1325--1330.
However, the length of the current light curve used in our analysis is $\sim 1770$ days, which only includes two complete cycles for a periodicity of  $\sim 625$ days.
Given the scattering of the data and the total length of the observations, the current data set is not suitable for the HHT.
Since there is an obvious gap without any data points at days 845--1085, we divided the data into two groups to inspect whether there is any periodic signal in a specific time interval.
These results are also presented in Fig.~\ref{DS_LSP}. 
Because the WWZ can also be employed to unevenly sampling data, we double checked those signals resolved from the DPS and from the LSP with the sub-divided data.
In the WWZ periodogram of Fig.~\ref{WWZ}, there is a strong signal at low frequencies ($\sim 625$ days) persisting across the whole data set.
The significance of this signal slightly decreases after day 1050, and a similar phenomenon can also be observed in Fig.~\ref{DPS}.
We cannot confirm such a long-term periodicity with the current data with limited cycles since the resolution of signals obtained from the WWZ is not good enough. 
Furthermore, no adequate conclusion can be made on this feature from DPS because the adopted window time (500 days) is shorter than the periodicity we are discussing.

\subsection{M81 X--6} 
We show the \swift\ long-term light curve of M81 X--6 in Fig.~\ref{LC}.
The LSPs obtained from both the unbinned data and 5-day average data are presented in Fig.~\ref{LSP}. 
The source varies by a factor of $\sim 4$ on a time scale of months. 
The LSP indicates that the strongest candidate periods are at about 110 days and 370 days, but they are not statistically significant ($< 99$\% red noise confidence level) when we take into account the entire data sets.

\begin{figure*} \centering
\psfig{file=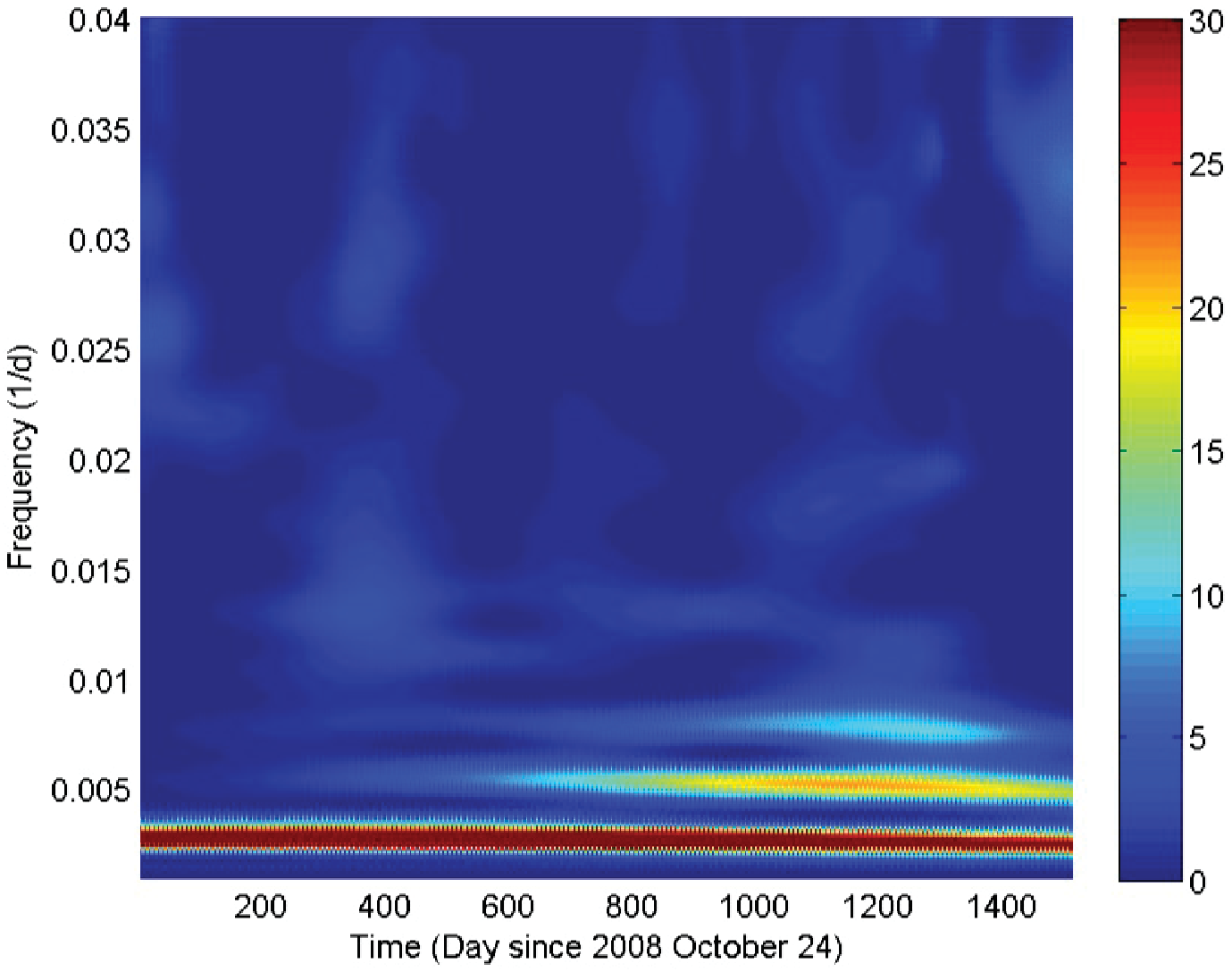,width=3.3in}
\psfig{file=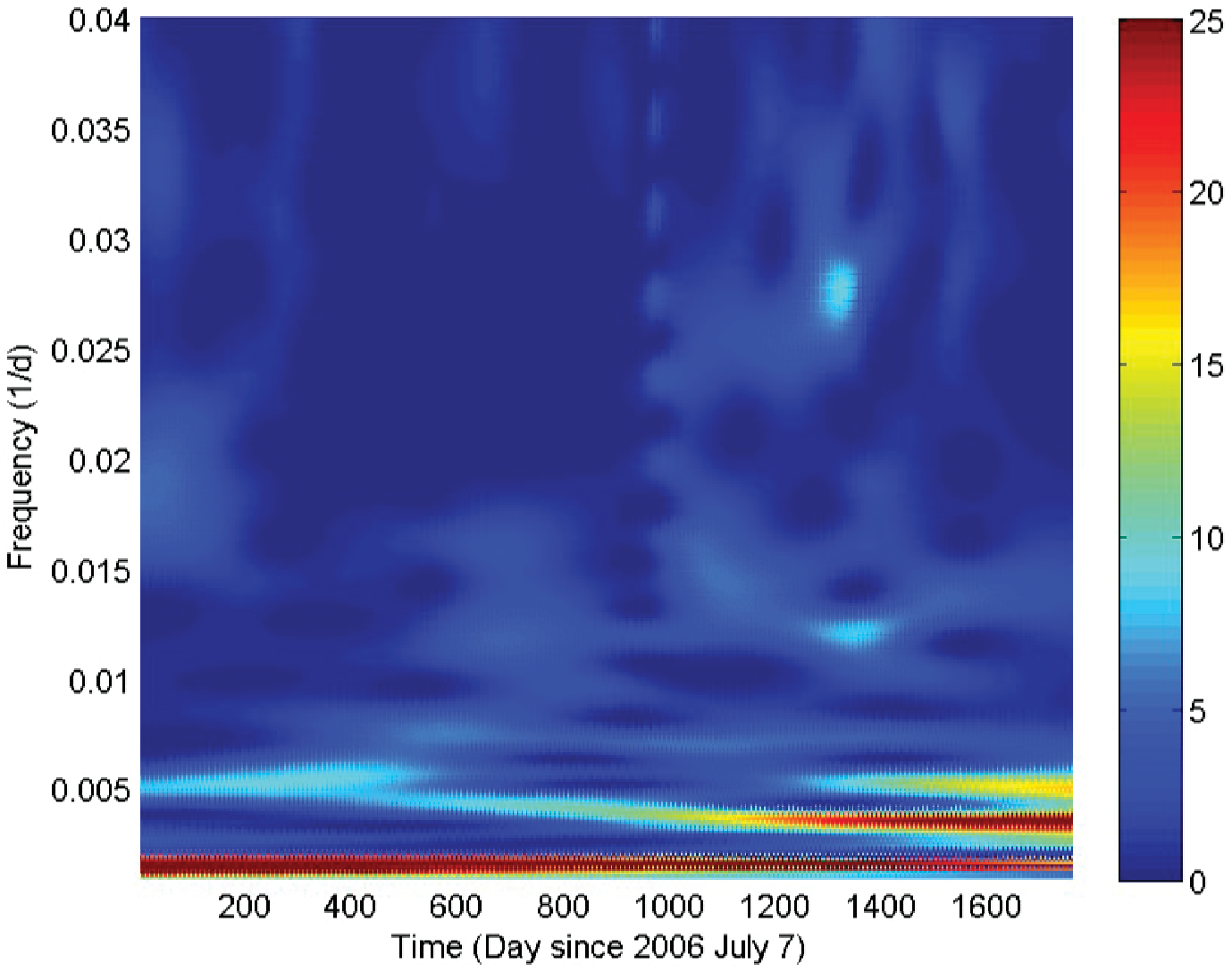,width=3.3in}
\psfig{file=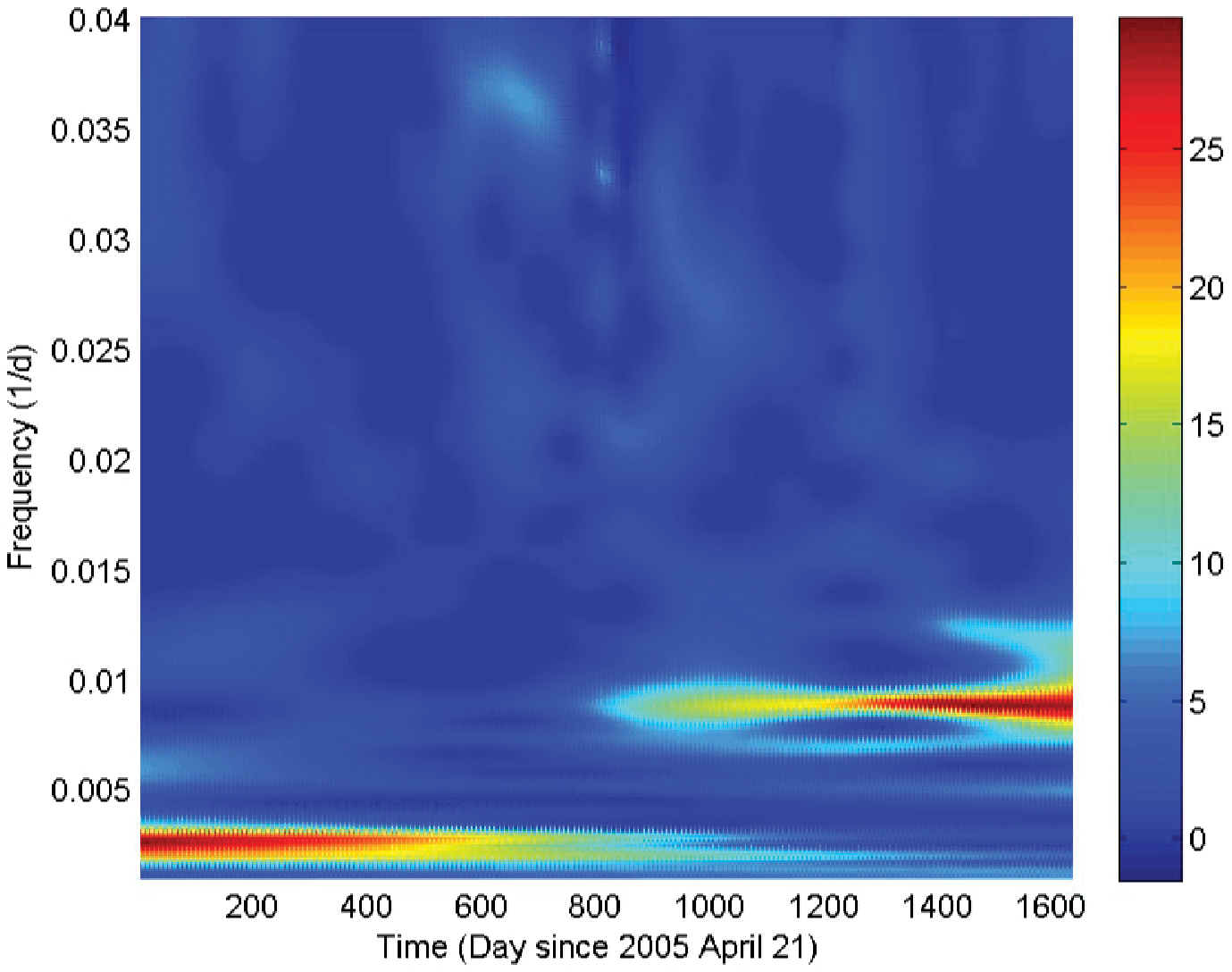,width=3.3in}
\psfig{file=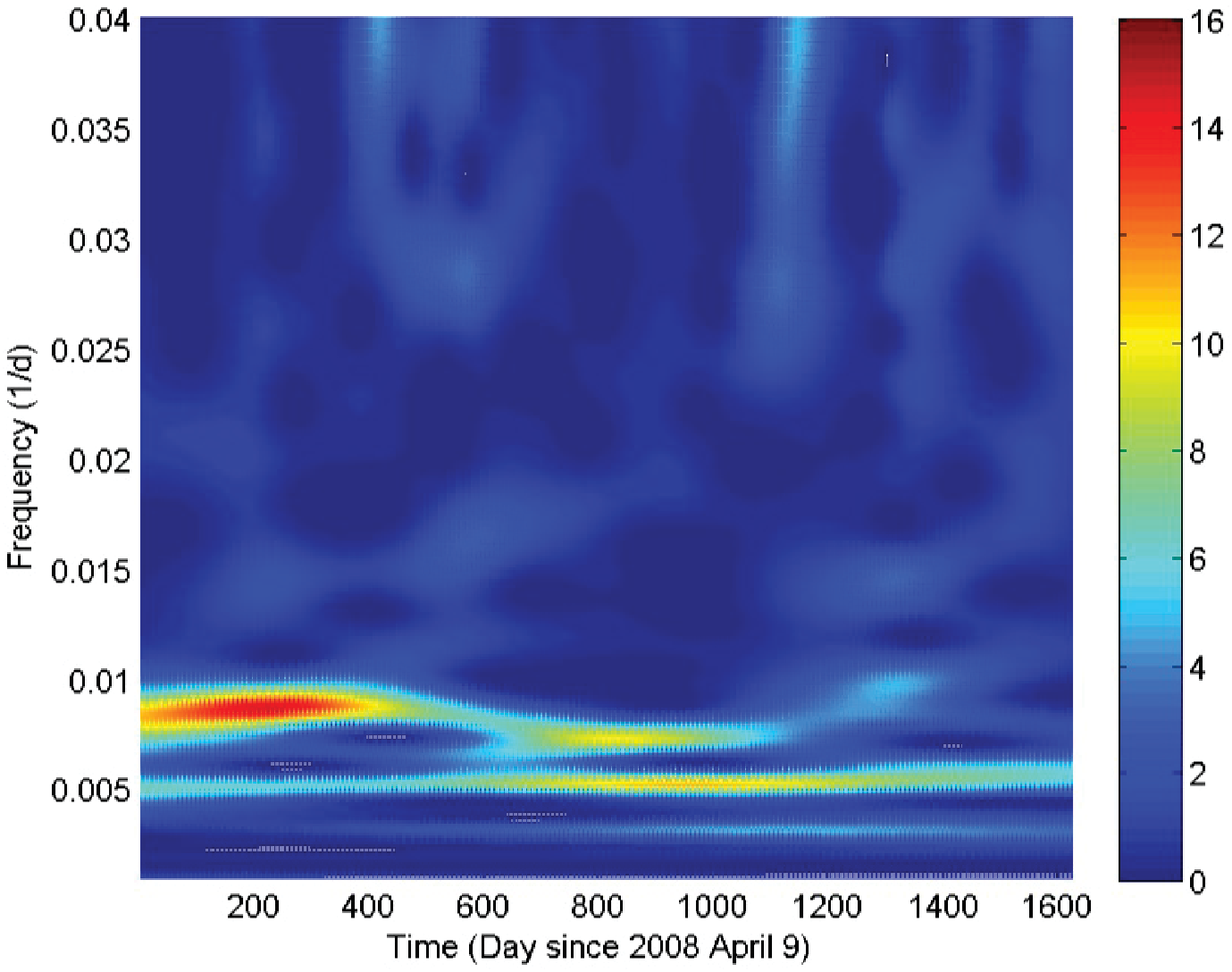,width=3.3in}
\caption{WWZ periodograms of \eso\ (upper left), \ho\ (upper right), M81 X--6 (lower left), and \ngc\ (lower right). 
All the periodograms were obtained with the WWZ stated in \S 3.2 using the re-binned data. The colour of the contour represents the strength of the significance defined through eq.~\ref{eqno7}.} 
\label{WWZ}
\end{figure*}

Only two signals are more significant than the 99\% white noise confidence level in the DPS (Fig.~\ref{DPS}), and this is consistent with the LSP in Fig.~\ref{DS_LSP}. 
The black curve describes the LSP obtained from the re-binned data sets of $< \sim$ day 740 and the red one represents the LSP obtained from the re-binned data sets of $\sim$ day 965--1645.
The period at $\sim$ 370 days is only prominent before day 740 while the period at $\sim$ 110 days is only significant after day 965. 
There are two obvious gaps without any data points in the light curve after day 965.
The length of the two gaps (at days 1160--1315 and 1350--1490) is 155 days and 140 days.
In addition, we also found that the duration with frequent cadence at days 965--1160 and 1490--1645 is 195 days and 155 days, respectively.
The duration of the gaps and the data with frequent cadence also propagate the fake signals in LSP of Fig.~\ref{LSP}. 
When we examined the signals with the WWZ shown in Fig.~\ref{WWZ}, the distinction between the two major signals and others are much more evident. 
These findings strongly suggest that those sub-signals found in the LSP of Fig.~\ref{LSP} are artificial effects.

\subsection{NGC\,5408 X--1} 

With \swift/XRT data covering about 1500 days, we plot the long-term light curve in Fig~\ref{LC}.   
We performed a power spectral analysis with LSP and the significance levels of white and red noises in the resulting power spectrum are shown in Fig.~\ref{LSP}. 
We do not find any significant periodicity; the highest peaks obtained from the original data and the re-binned data sets (5 days per bin) are at $112.1 \pm 1.0$ days and $189 \pm 3$ days, respectively. 
These results are consistent with the proposed $(112.6 \pm 4)$-day quasi-periodic modulation obtained by \citet{Pasham2013} and the $\sim$ 187-day periodicity detected by \citet{Grise2013}.
Because there is a large deviation between the obtained peak at $\sim 190$ days and the proposed orbital period ($243\pm 23$ days) of this system \citep{Pasham2013}, it is difficult to establish a direct connection between our results and the suggested orbital modulation.

In the DPS (Fig.~\ref{DPS}), it is clear that the signal near 112 days is quite strong during the first $\sim 500$ days of observations, but it has become weaker after that. 
The quasi-periodic signal near 112 days has significantly decreased after day 800 and evolved into a quasi-periodicity of $\sim$ 190 days.
On the other hand, the $\sim$ 190-day period has been evident after day 800 and was quite weak before day 500.
In order to clarify this, we used the LSP obtained in different time intervals to determine the strength of signals obtained in different stages and to avoid any contaminations from artificial signals arisen from the length of empty observations.
The black and red curves in Fig.~\ref{DS_LSP} represent the LSP resulted from the re-binned data of $\sim$ day 0--535 and $\sim$ day 0--1230 since MJD 54565.8, respectively.
An exchange of the significance between two major signals ($\sim$ 112 days and 190 days) is clearly shown although they are not statistically significant, and the similar phenomenon can also be verified by the WWZ shown in Fig.~\ref{WWZ}.

\begin{figure*} \centering
\psfig{file=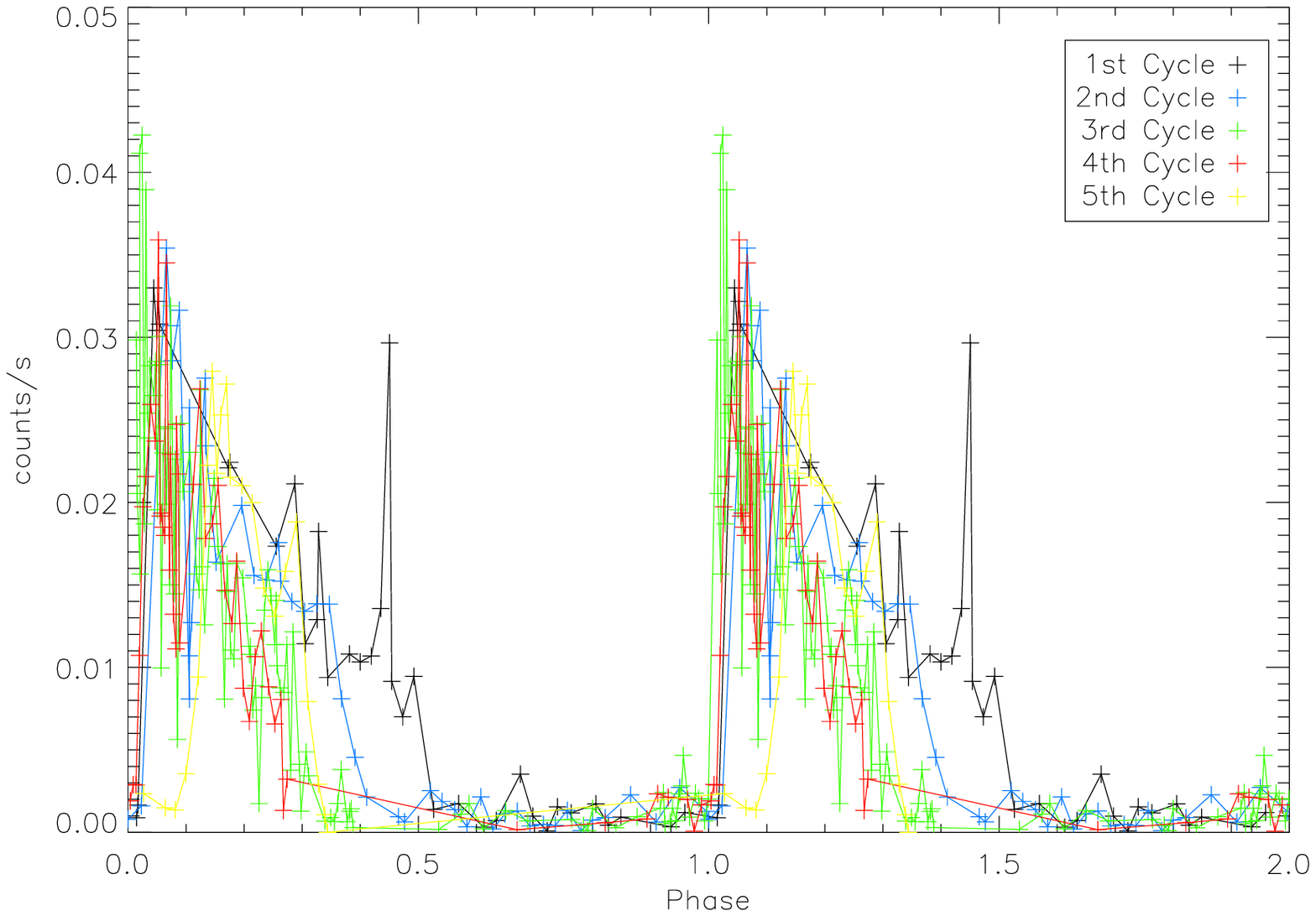,width=6in, height=3.3in}
\psfig{file=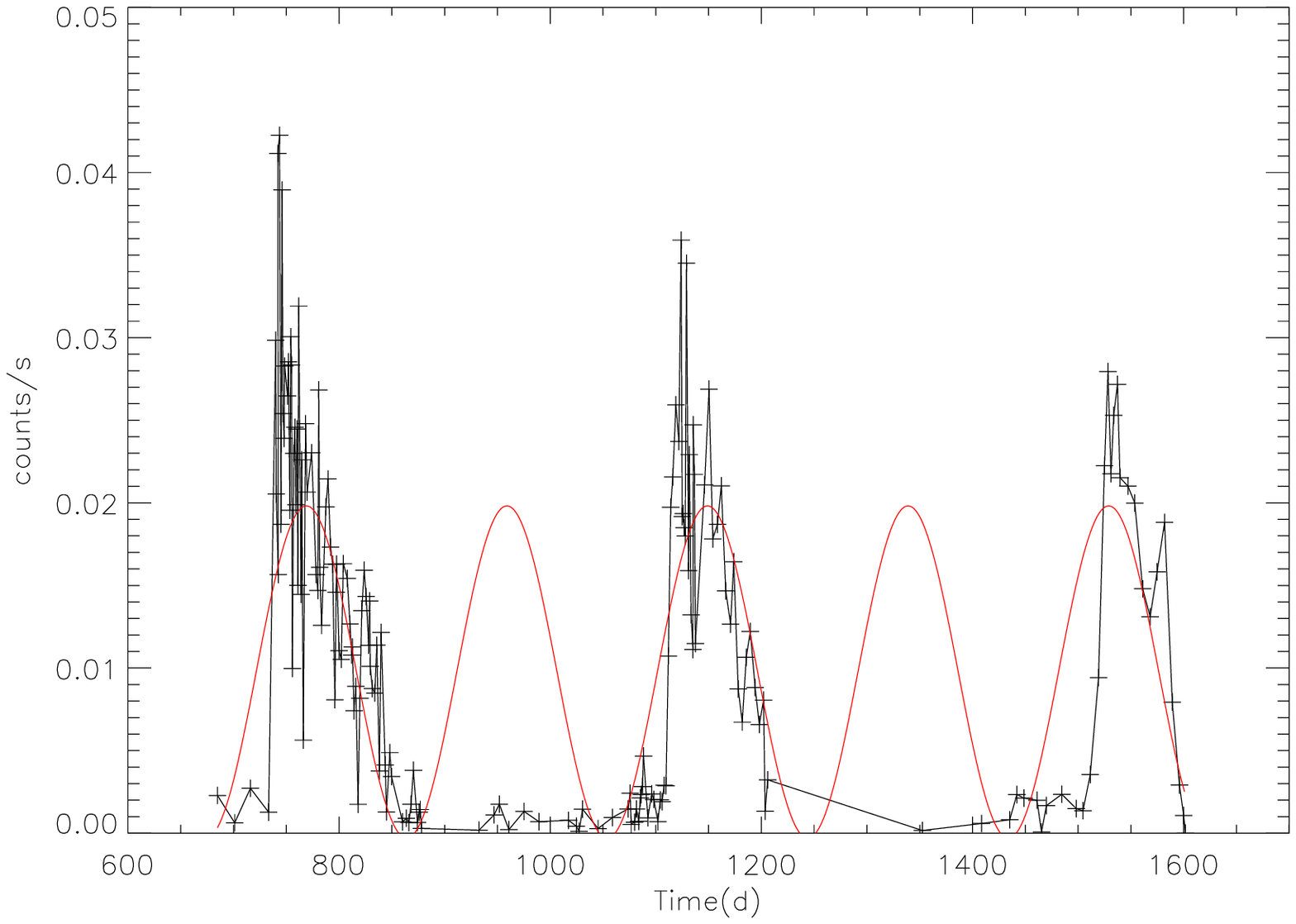,width=6in, height=3.3in}
\caption{Folded light curve and sinusoidal fit to the sub-signal obtained from \eso. {\it Top}: The unbinned light curve was folded with 370 days to demonstrate the recurrence of each outburst. The phase zero was set according to the first data point used in our timing analysis. This light curve includes 5 outbursts recorded by \swift. The recent outburst has a significant delay, which can be clearly seen from the rising phase of main peak. {\it Bottom}: The light curve taken from the third to the last outbursts overlapped with a sub-signal of 190 days. This signal provides a good fit to the duty cycle of the third and the fourth outbursts as shown in Fig.~\ref{LSP} and Fig.~\ref{DS_LSP}.}
\label{FLC}
\end{figure*}

\section{Discussion}

We generated LSP and DPS for four ULXs from the long-term monitoring \swift\ XRT data, with the aim of revealing periodic signals embedded in the light curves. 
The results were also examined with the WWZ.
In the case of \eso, we also used HHT to further study its periodicity.
Here, we discuss the results for each object.

\subsection{\eso}

Both the LSP and DPS shown in Fig.~\ref{LSP} and Fig.~\ref{DPS} present a strong persistent periodic signal at $\sim 380$ days.
This detection is consistent with the mean separation between the start of different outbursts or between their peak luminosities ($\sim$ 350, 373, 375 and 407 days).
The gradual delay of each outburst also reflects on the result of DPS, and a clear bifurcation from the main signal can also be seen.
Because the biforked feature is only significant in the data $>$ day 720, we divided the data sets observed with similar cadence ($< \sim 1210$ days) in two segments that both include the entire duty cycle of three outbursts to trace the variation of the periodic signals.
The LSP in Fig.~\ref{DS_LSP} yielded from the data which include outbursts from the second to the fourth reveals some other quasi-periodic signals that correspond to the harmonics of the main signal at $\sim 380$ days.
If we reduce the length of the light curve to include the third and the fourth outbursts only, we find that the significance of these signals is enhanced.       

The Hilbert spectrum and the WWZ provide us an opportunity to verify the results yielded from LSP and DPS.
We compared the DPS and the Hilbert spectrum in Fig.~\ref{HHT_LSP}.
The contour represents the power of DPS generated with a moving window of 500 days through the re-binned light curve. 
The colour map represents the Hilbert energy defined as the square of the amplitude, and it was smoothed using a Gaussian filter.
It is clear that the Hilbert spectrum yielded from the normalised HHT passes through the peak of DPS, indicating that the result of the Hilbert spectrum and the DPS are consistent. Furthermore, the Hilbert spectrum provides a much higher/better resolution to investigate the structures of the variability/instability caused by the intra-wave modulation, i.e. the frequency modulation within one cycle.  It is caused by the non-sinusoidal nature of the modulation because the modulation shape contains a sharp rise and an exponential decay. 
The maximum frequency of the the main signal due to the intra-wave modulation appears at $\sim$ 390, 745 and 1125 days (Fig.~\ref{HHT_LSP}), which are similar to the epochs of the peak luminosity of each outburst.

Comparing to the DPS in Fig.~\ref{DPS}, we do not find a biforked feature on the WWZ periodogram.
The reason is that the Lomb-Scargle method tends to describe the data with a smooth sinusoidal wave characterized by a longer period when the moving window approaches the flat low count rate state. 
Nevertheless, the gradual delay of the major signal can still be seen in Fig.~\ref{WWZ} if we zoom in the frequency domain. 
We can also find that a new signal begins to appear after day 510, and its power is enhanced after day 760.
Because day 510 is at the quiescent state after the second outburst and day 760 is close to the peak of the third outburst, we speculate that the emergence and the enhancement of this new signal might be due to a change of the shape of the light curve in outbursts.
The best fit to the cycle length of the first four complete outbursts can be estimated as $\sim$ 370 days from the strongest signal in the LSP.    
We then folded the light curve with this periodicity to examine the difference of the outburst light curve profiles in each cycle.
As shown in the top panel of Fig.~\ref{FLC}, we use different colours to mark each cycle.
The black curve characterizes the first cycle, while the blue one represents the second cycle. 
The third and the fourth outbursts are depicted with green and red colours, repectively. 
The most recent outburst denoted by the yellow curve recurred with a relatively long separation ($\sim 407$ days) and clearly shows a delay during its rising phase.
The first four outbursts start with a sharp rise near phase 0, but the profiles of their decays are different. 
The duty cycle of the outburst decreases from $\sim 50$\% in first cycle to $\sim 30$\% in the last cycle.
The third and the fourth outbursts have similar phase for each peak, and a good fit to their duty cycle can be modelled by a consecutive sinusoidal signal with a periodicity of 190 days as shown in the bottom panel of Fig.~\ref{FLC}.
This periodicity is consistent with the signal ($190\pm 3$ days) that we obtained in Fig.~\ref{LSP}.

A third signal of $\sim 125$ days in Fig.~\ref{DS_LSP} is roughly consistent with the length of the duty cycle of recent outbursts.
These results give an interpretation to the bifurcation of the main signal shown in the DPS and the emergence of all the sub-signals shown in the DPS and the WWZ, and these quasi-periodicities are indeed originated from the harmonics of the main signal.      
Because the distribution of the data is highly non-linear and the duty cycle of the third and fourth outburst is much shorter, the EEMD tends to decompose the peak/amplitude into another IMF and this feature can also be seen in the Hilbert spectrum. 
Therefore the power/energy of the main signal during the first two outbursts shown in the DPS is weaker than that during the next two outbursts (Fig.~\ref{DPS}) while the Hilbert spectrum presents an opposite result. 
Moreover, we see an obvious signal varying between 125 and 250 days at day 510--1310 in the Hilbert spectrum as well. 

Since the main periodicity embedded in the light curve of \eso\ arises from the separation of each outburst, this signal is not stable because of the delay between outbursts. 
We can clearly see this feature in Fig.~\ref{DPS}, and the main period increases to $\sim 380$ days if we take into account the latest outburst.  
To explain the $\sim 380$-day modulation, it has been suggested that the companion star is in a highly eccentric orbit with the central black hole \citep{Lasota2011,Soria2013}. 
Our analyses, however, indicate that the modulation is not stable. 
More recently, \citet{Godet2014} proposed that the orbital period could change between different orbits as a consequence of stochastic fluctuations.
As an alternative explanation we speculate that this modulation represents a superorbital periodicity, while the orbital period of the system is likely $< 125$ days \citep{Kotze2012}. 
\citet{KL2014} suggest that the system may be similar to the Galactic microquasar SS433 for which the X-ray modulation is due to precession of an X-ray beam.

\begin{figure*} \centering
\psfig{file=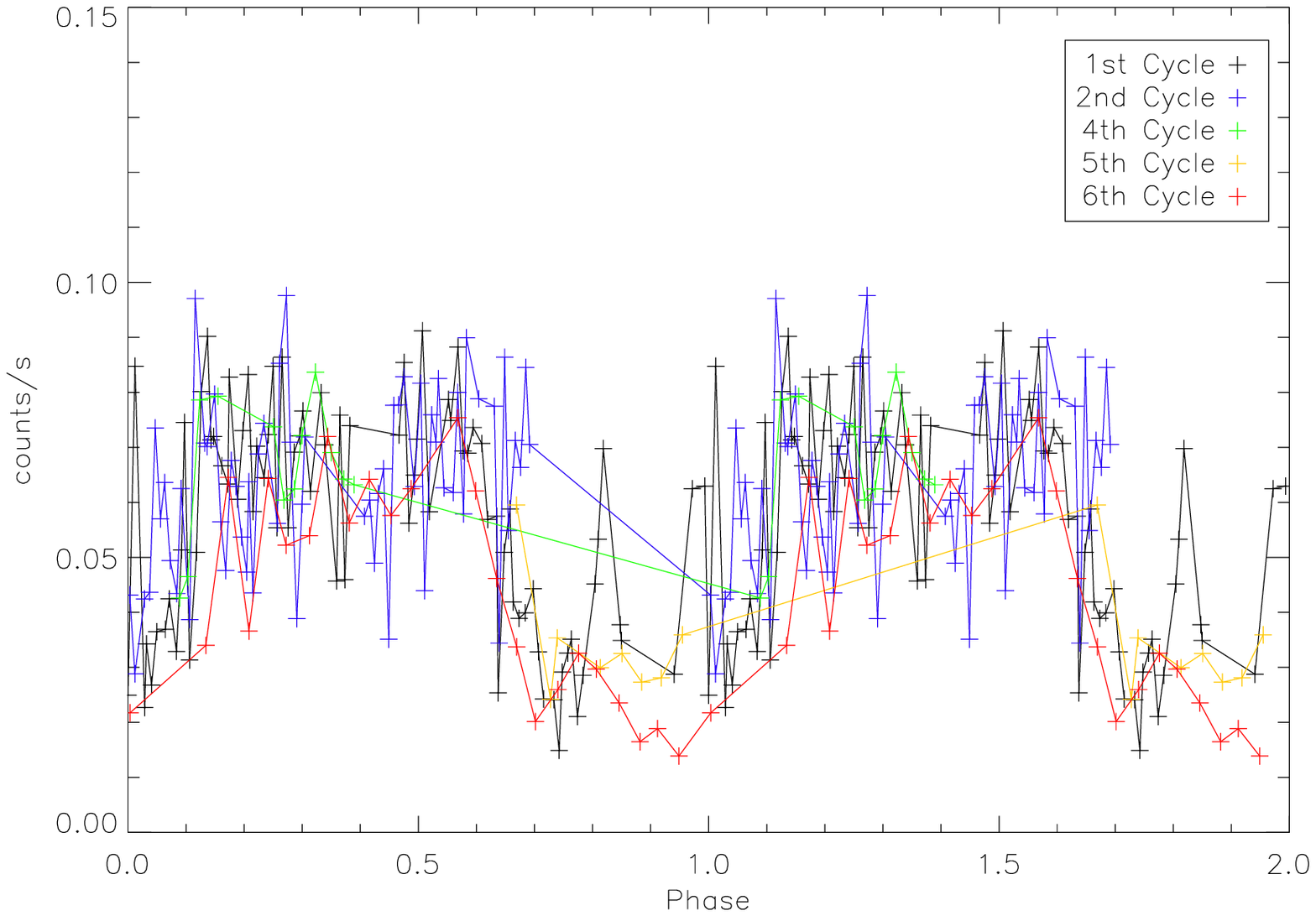,width=6in, height=3.3in}
\psfig{file=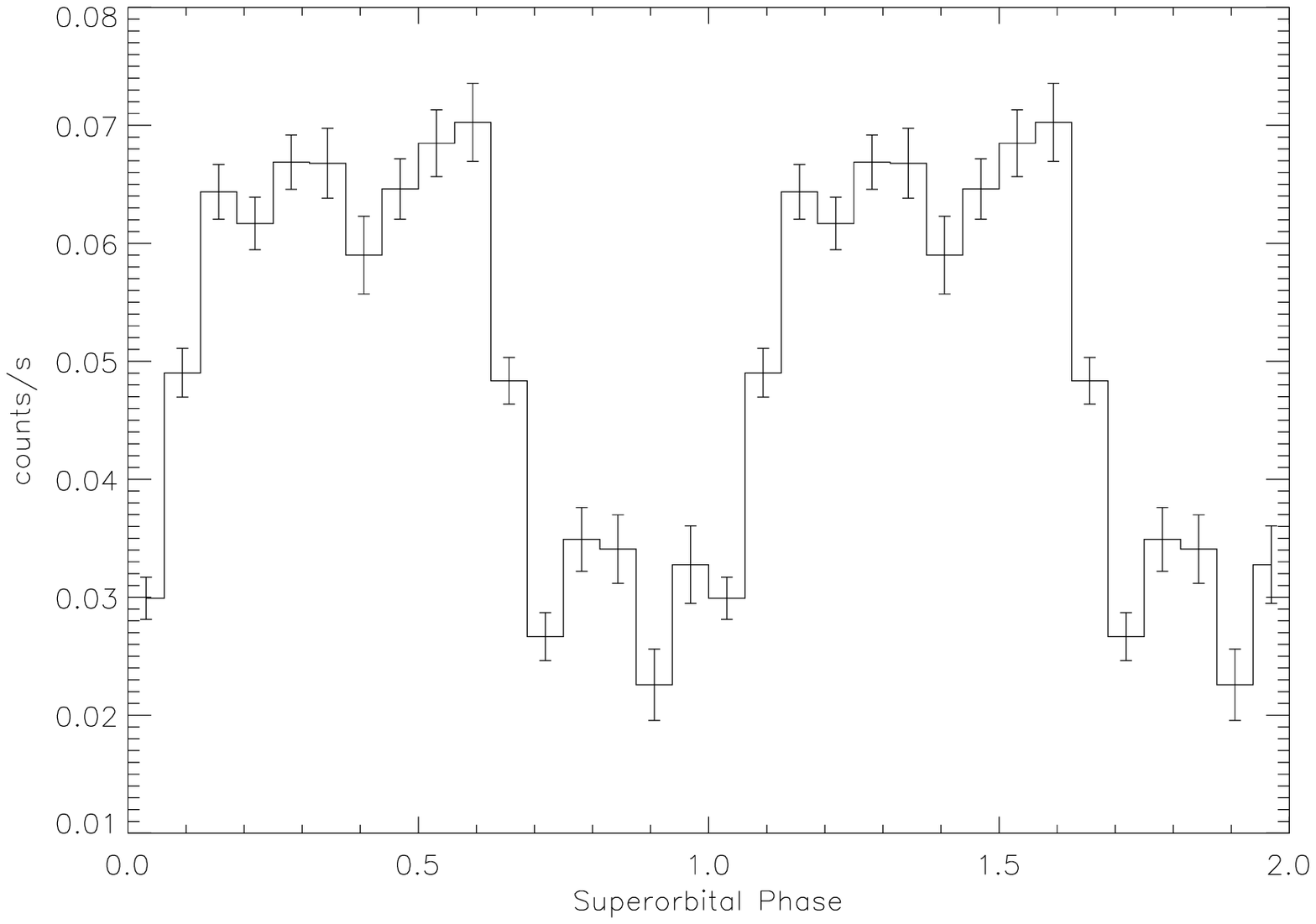,width=6in, height=3.3in}
\caption{Folded light curve obtained from the \swift\ data of M81 X--6 between day $\sim 965-1645$. {\it Top}: The unbinned light curve was folded with 112.4 days. The phase zero (at $\sim$ day 966.5) was set at the epoch of the first observation in the second segment of the light curve. Different colours are used to show the data of each cycle. We ignored the 3rd cycle without any observation and the final cycle with only 2 data points. {\it Bottom}: Binned light curve of the last segment folded with the 112.4-day quasi-periodic signal.} 
\label{M81_FLC}
\end{figure*}

\subsection{\ho}

\citet{Kaaret2009b} detected two strong possible signals near 28.8 and 64.5 days with the \swift\ data taken between 2006 July and 2009 June.
However, both signals are not statistically significant when the 11 observations before MJD 54200 are removed and only 177 observations after MJD 54801 (day 878 in Fig.~\ref{LC}) are taken into the consideration \citep{Kong2010}.
Here, we have more observations covering $\sim 1770$ days to investigate the long-term periodic signal of \ho. 
By taking red noise into account, the two periods are not significant.

We also find a prominent signal corresponding to a quasi-periodicity of $\sim 625$ days ($\sim 0.016$~1/day in Fig.~\ref{LSP}).
We find that this signal roughly corresponds to the separation between each peak or dip event although some of them seem to appear randomly.
According to Fig.~\ref{DPS} and Fig.~\ref{DS_LSP}, the signal of $\sim 625$ days is much more prominent in the beginning (before day $\sim 845$) of the re-binned light curve, but the length of this segment is shorter than two complete cycles.
Although this long-term signal has a stronger significance in the WWZ (see Fig.~\ref{WWZ}), the length of the entire light curve we used in the analysis is too short (only about 2.8 cycles) to confirm this periodicity.  
In addition, we must emphasize that the window size adopted in the current DPS is only 500 days and it is also the reason why we only show the signals of interest with periodicity $< 500$ days in most LSPs.
All the above reasons make the confirmation of such a long periodicity difficult.
Because the light curve of \ho\ is quite spiky, it is not appropriate to interpolate the data sets to generate an evenly sampled light curve for HHT. 
In order to confirm the long-term periodicity of \ho, more observations with similar cadence are required in the future.   

\subsection{M81 X--6}

For M81 X--6, except for a major gap with only two observations at day $\sim 740-965$ and two other gaps with no observations at $\sim 1160-1315$ and 1350--1490, the mean cadence of the data set is 2.8 days. 
Therefore, the \swift\ light curve is not ideal to examine the proposed 1.8-day orbital period \citep{Liu2002}, but it provides us an opportunity to study the long-term variability of this high-mass X-ray binary system.     
Although there is no significant period in the LSP (Fig.~\ref{LSP}), there are two significant periods when we consider the light curves in different epochs (Fig.~\ref{DS_LSP}).
From the DPS (Fig.~\ref{DPS}) and the LSP obtained from different stages shown in Fig.~\ref{DS_LSP}, we find that the $\sim 370$-day period is only prominent in the first segment ($< \sim$ day 740) of the light curve while the $\sim 110$-day one is only prominent in the last segment ($\sim$ day 965-1645) of the light curve.
The WWZ periodogram also clearly demonstrates two strong quasi-periodicities existed in a different time interval.
Comparing with the DPS, WWZ is unaffected by the noise generated from the window effect and all spurious sub-signals are less significant in the WWZ map.   

In the first segment of the light curve, the dip events ($< 0.015$ cts/s) take place at days $\sim 170$ and $\sim 555$ and the peak events ($> 0.078$ cts/s) appear at days $\sim 25$ and $\sim 380$. 
The separation between the dip or the peak luminosity in the light curve provides a possible explanation to the quasi-periodic signal at $\sim 370$ days.
Because the length of the light curve in the first segment is only about 740 days, which only covers about two cycles of the $\sim 370$-day quasi-periodic signal, it is difficult to confirm the stability of this signal with this limited data set.     

The $\sim$110-day quasi-periodic signal is not only the most significant one we obtained in the second segment of the light curve, but also the most significant one in the unbinned data.
The light curve folded with $\sim 110$ days is shown in Fig.~\ref{M81_FLC}.
Although there are two obvious gaps in the 3rd cycle and between the 4th and the 5th cycles, observations in the last segment folded with the $\sim 110$-day quasi-periodic signal show similar structure as demonstrated in the binned folded light curve. 
According to the DPS and the WWZ, this signal is still significant from more recent observations.
Future observations with similar cadence are required to infer the nature of this quasi-periodicity.   

\subsection{\ngc}
With a long-term monitoring of \ngc, we find that the most obvious signals from LSP and DPS are $\sim 112$ days and $\sim 190$ days.
However, none of them are statistically significant, and this result is consistent with \citet{Grise2013}.
In the analysis with DPS, the $\sim 112$-day period is stronger during the first $\sim 500$ days of observations, which is consistent with the $(115.5\pm 4)$-day period detected by \citet{Strohmayer2009a} using the data of the first $\sim 485$ days.
Another quasi-periodicity of $\sim 190$-day starts to appear after day 400 and becomes prominent after day 800.  

We do not find any strong signal with a period longer than 200 days to support a proposed orbital period of $243\pm 23$ days \citep{Pasham2013}.
Another weak signal of $\sim 310$ days arisen at day 600 shown in Fig.~\ref{LSP} and Fig.~\ref{WWZ} is not related to the proposed orbital period as well, even if we take into account its 95\% uncertainty.
Since the recurrence of the X-ray dips related to the claimed periodicity of $\sim 243$ days \citep{Pasham2013} only occurs before day 1200 and the time duration of each cycle is quite unstable (251 days, 211 days, 276 days and 235 days), this signal is unlikely real and future monitoring observations may see a suppression of the signal.   
\citet{Grise2013} also indicated that the dip events do not repeat at every cycle.
Such phenomenon can also be seen in some Galactic transient sources (e.g., GRO J1655--40; \citealt{Kuulkers2000}).
Because the transient behaviour can cause a change on the structure of the accretion disc, 
the efficiency in X-ray irradiation of the disc will change accordingly causing different behaviours of dipping activities \citep{Kuulkers2000}.
However, \ngc\ is a persistent source without any obvious change in X-rays on timescale of years. A lack of dipping activity in some epochs has been seen in Galactic X-ray dipping sources (e.g., X1916-053 and XB 1254-690, see \citealt{Chou2001,Homer2001} and \citealt{Smale99}) and it may be due to a decrease of the vertical structure of the accretion disc.
Alternatively, a tilted and precessing disc may provide a possible scenario on the variability of the periodicities and the disappearance of dips \citep{Diaz2009}.  
The presence of two relatively strong quasi-periodic signals ($\sim 112$ days and $\sim 190$ days) in the light curve may still have certain relation with the orbital period of \ngc. 
Such a feature can be caused by variations of the accretion rate due to a precession of the accretion disc induced by the tidal force \citep{Kotze2012}, or instabilities of disc warping driven by irradiation \citep{Charles2008,Ogilvie2001}.
Last but not least, the quasi-periodic signals may be associated with a precessing jet \citep{Foster2010}.
We therefore suggest the detected periods to be superorbital periods since these signals are only prominent in a specific time interval of the long-term light curve although it is not entirely clear which mechanism induces these signals.
According to previous investigations between superorbital periodicities and orbital periods of Galactic X-ray binaries \citep{Kotze2012}, we support the conclusion given by \citet{Grise2013}, who report an orbital period of $< 40$ days.

\subsection{Concluding Remarks}

We have used the \swift\ archive to investigate the long-term X-ray variability of four ULXs.
With timing analyses through the LSP, DPS and WWZ, we detect the following intriguing periodic signals embedded in their light curves: $\sim 380$ days for \eso, and $\sim 370$ days and $\sim 110$ days for M81~X--6.
A brief summary of all potential periods is shown in Table~\ref{period}.

Except for the main signal detected from \eso, all other period candidates are only prominent in a specific time interval.
The $\sim 380$-day periodic signal of \eso\ corresponds to the separation between outbursts, and it was claimed to represent an orbital modulation \citep{Lasota2011}.
Among the four ULXs discussed in this paper, only \eso\ has a data set characterized by regular monitoring, which is adequate to be examined with the HHT.  
This is also the first time that HHT has been applied to study long-term X-ray variability of a ULX.
As compared with the DPS, the WWZ and the HHT are independent of the window time, and the Hilbert spectrum provides a better resolution to discriminate the evolution of each signal.  
We find that the separation between each outburst varies over time, and suggests this main signal to be a superorbital period. 
This may also support the idea of a precessing X-ray beam \citep{KL2014} although a changing orbital period could also be possible \citep{Godet2014}. At the time of writing, the new outburst has started in 2015 early-January, more than 460 days since the last outburst \citep{Kong2015}. More monitoring observations in the future are required to distinguish among different models.

\begin{table} 
\caption{Summary of our detected signals significant at $\gtrsim 99$\% against white and red noise}\label{period}
\begin{tabular}{ccc} 
\hline \hline Source & Period/MJD range & Possible Origin 
\\ \hline
ESO 243--49 & $\sim 380$ days & orbital period?
\\ HLX--1 &  ($\sim$MJD 55050-56561) &    superorbital period?
\\ \hline
Holmberg IX & $\sim 625$ days$^{*}$ ? & \multirow{2}{*}{superorbital period ?}
\\ X--1 & ($\sim$MJD 54803.7-57439.4) & 
\\ \hline
\multirow{4}{*}{M81 X--6} & $\sim 370$ days & \multirow{2}{*}{superorbital period ?}
\\  & ($\sim$MJD 54924.6-56554.7) & 
\\  & $\sim 110$ days & \multirow{2}{*}{superorbital period ?}
\\  & ($\sim$MJD 55889.6-56554.7) & 
\\ \hline 
\ngc\ & None & ---
\\ 
\hline 
\hline 
\multicolumn{3}{c}{$^{*}$ Too long to be definitely confirmed with current investigations}
\end{tabular}
\end{table}

The period candidates of other three ULXs may range from $\sim 100$ days to $\sim 600$ days. Apart from noise and artifacts, all the candidate periods are only significant in a specific epoch. This suggests that they are not associated with any stable mechanism such as orbital motion. Instead, such long-term ($>$ 100 days) X-ray quasi-periodic variations are likely related to superorbital periods that are thought to be due to radiation-driven warping of accretion discs \citep{Ogilvie2001} or tidal interaction-induced disc precession \citep{WK91}. 
Alternatively, mass transfer rate-related events such as X-ray state changes and disc instability can also cause long-term modulations \citep{Kotze2012}. In particular, there are two intermittent quasi-periodicities for both \ngc\ and M81 X--6, suggesting that the quasi-periods are changing or evolving. They resemble some Galactic X-ray binaries that show similar behaviour (e.g., Cyg X--2 and SMC X--1; \citealt{Kotze2012}) and it has been suggested that a warped disc could lead to an unstable steadily precessing disc, causing quasi-periodic behaviour \citep{Ogilvie2001}. 
We note that there are many uncertainties on the physical parameters of ULXs. 
To determine the origin of superorbital periods of ULXs, one has to know at least the mass ratio between the companion and the compact star ($q=M_C/M_X$) and the binary separation. Unfortunately, it is very difficult to get these parameters for ULXs.
For the three ULXs discussed here (i.e., excluding \eso), only M81 X--6 has better constraints on the black hole mass and the nature of the companion. The masses of the black hole and companion star are estimated ($M_{X}=18M_\odot$, $M_C=23M_\odot$) such that $q$ can be derived. 
In this case, we can rule out a tidal interaction-induced disc precession scenario that requires $q<0.25-0.33$ \citep{WK91}. For a warped disc, the binary separation and the mass ratio suggest that M81 X--6 lies in the intermediate instability zone for radiation-driven warping in X-ray binaries (see Figure 1 of \citealt{Kotze2012}). 
The quasi-periodic variability may represent the switching timescale between a warped disc and a flat disc.

If all of these intriguing signals do represent superorbital periods, the real orbital periodicities of these systems are expected to be much shorter \citep{Kotze2012}. This will give us constraints on the binary separation as well as the black hole mass.  
In order to understand the physical nature of the quasi-periodic modulations of the four ULXs discussed here, a regular long-term X-ray monitoring is required in the future. 
More importantly, we demonstrate that dynamic timing analysis is important to investigate the long-term X-ray behaviour of ULXs especially when the periodic behaviours are intermittent or varying. 

\section*{Acknowledgments} 
We would like to thank the anonymous reviewer for his/her thorough review and constructive comments to improve the paper.
This work made use of data supplied by the UK Swift Science Data Centre at the University of Leicester. 
This project is partially supported by the Ministry of Science and Technology of the Republic of China (Taiwan) through grant NSC~101-2112-M-039-001-MY3.
C.~P.~H. and Y.~C. are supported by the Ministry of Science and Technology of Taiwan through the grant NSC~102-1221-M-008-020-MY3.
A.~K.~H.~K. is supported by the Ministry of Science and Technology of Taiwan through grant 103-2628-M-007-003-MY3. 
D.~C.~C.~Y. gets the financial support from the Ministry of Science and Technology of Taiwan through NSC~101-2115-M-030-004.

\label{lastpage}

\end{document}